\documentclass[fleqn,usenatbib]{mnras}

\usepackage{newtxtext,newtxmath}
\usepackage[T1]{fontenc}

\DeclareRobustCommand{\VAN}[3]{#2}
\let\VANthebibliography\thebibliography
\def\thebibliography{\DeclareRobustCommand{\VAN}[3]{##3}\VANthebibliography}

\usepackage{graphicx}	% Including figure files
\usepackage{amsmath}	% Advanced maths commands
\newcommand{\kms}{km s$^{-1}$}
\title[Negative and Positive Feedback from IC443]{Negative and Positive Feedback from a Supernova Remnant with SHREC: A detailed Study of the Shocked Gas in IC443}

\author[G. Cosentino et al.]
{G. Cosentino$^{1}$\thanks{E-mail:giuliana.cosentino@chalmers.se},
I. Jim\'{e}nez-Serra$^{2}$,
J. C. Tan$^{1,3}$, 
J. D. Henshaw$^{4}$,
A. T. Barnes$^{5}$,
C.-Y. Law$^{1,6}$,
S. Zeng$^{7}$,
\newauthor
F. Fontani$^{8}$
P. Caselli$^{9}$,
S. Viti$^{10,11}$,
S. Zahorecz$^{12,13}$
F. Rico-Villas$^{2}$,
A. Meg\'ias$^{2}$,
M. Miceli$^{14,15}$,
S. Orlando$^{15}$,
\newauthor
S. Ustamujic$^{15}$,
E. Greco$^{15,16,17}$,
G. Peres$^{14,15}$,
F. Bocchino$^{15}$,
R. Fedriani$^{1}$,
P. Gorai$^{1}$,
L. Testi$^{6}$,
J. Mart\'{i}n-Pintado$^{2}$\\
$^{1}$Space, Earth and Environment Department, Chalmers University of Technology, SE-412 96 Gothenburg, Sweden\\
$^{2}$Centro de Astrobiolog\'{i}a (CSIC/INTA), Ctra. de Torrej\'on a Ajalvir km 4, Madrid, Spain\\
$^{3}$Department of Astronomy, University of Virginia, 530 McCormick Road
Charlottesville, 22904-4325 USA\\
$^{4}$Max Planck Institute for Astronomy, K\"onigstuhl 17, D-69117 Heidelberg, Germany\\
$^{5}$Argelander-Institut f\"ur Astronomie, Universit\"at Bonn, Auf dem H\"ugel 71, 53121, Bonn, Germany\\
$^{6}$European Southern Observatory, Karl-Schwarzschild-Strasse 2, D-85748 Garching, Germany\\
$^{7}$Star and Planet Formation Laboratory, Cluster for Pioneering Research, RIKEN, 2-1 Hirosawa, Wako, Saitama, 351-0198, Japan\\
$^{8}$INAF  Osservatorio Astronomico di Arcetri, Largo E. Fermi 5, 50125 Florence, Italy\\
$^{9}$Max Planck Institute for Extraterrestrial Physics, Giessenbachstrasse 1, 85748 Garching bei M\"unchen, Germany\\
$^{10}$Leiden Observatory, Leiden University, PO Box 9513, 2300 RA Leiden, The Netherlands\\
$^{11}$Department of Physics and Astronomy, University College London, Gower Street, London, WC1E 6BT, UK\\
$^{12}$Department of Physical Science, Graduate School of Science, Osaka Prefecture University, 1-1 Gakuen-cho, Naka-ku, Sakai, Osaka 599-8531, Japan\\
$^{13}$National Astronomical Observatory of Japan, National Institutes of Natural Science, 2-21-1 Osawa, Mitaka, Tokyo 181-8588, Japan\\
$^{14}$Dipartimento di Fisica e Chimica E. Segrè, Università di Palermo, Via Archirafi 36, 90123 Palermo, Italy\\
$^{15}$INAF-Osservatorio Astronomico di Palermo, Piazza del Parlamento 1, 90134 Palermo, Italy\\
$^{16}$Anton Pannekoek Institute for Astronomy, University of Amsterdam, Science Park 904, 1098 XH Amsterdam, The Netherlands\\
$^{17}$GRAPPA, University of Amsterdam, Science Park 904, 1098 XH Amsterdam, The Netherlands\\}
\date{Accepted XXX. Received YYY; in original form ZZZ}

\pubyear{2021}

\begin{document}
\label{firstpage}
\pagerange{\pageref{firstpage}--\pageref{lastpage}}
\maketitle

\begin{abstract}
Supernova remnants (SNRs) contribute to regulate the star formation efficiency and evolution of galaxies. As they expand into the interstellar medium (ISM), they transfer vast amounts of energy and momentum that displace, compress and heat the surrounding material. Despite the extensive work in galaxy evolution models, it remains to be observationally validated to what extent the molecular ISM is affected by the interaction with SNRs. We use the first results of the ESO-ARO Public Spectroscopic Survey SHREC, to investigate the shock interaction between the SNR IC443 and the nearby molecular clump G. We use high sensitivity SiO(2-1) and H$^{13}$CO$^+$(1-0) maps obtained by SHREC together with SiO(1-0) observations obtained with the 40m telescope at the Yebes Observatory. We find that the bulk of the SiO emission is arising from the ongoing shock interaction between IC443 and clump G. The shocked gas shows a well ordered kinematic structure, with velocities blue-shifted with respect to the central velocity of the SNR, similar to what observed toward other SNR-cloud interaction sites. The shock compression enhances the molecular gas density, n(H$_2$), up to $>$10$^5$ cm$^{-3}$, a factor of >10 higher than the ambient gas density and similar to values required to ignite star formation. Finally, we estimate that up to 50\% of the momentum injected by IC443 is transferred to the interacting molecular material. Therefore the molecular ISM may represent an important momentum carrier in sites of SNR-cloud interactions. 
\end{abstract}


%%%%%%%%%%%%%%%%%%%%%%%%%%%%%%%%%%%%%%%%%%%%%%%%%%

%%%%%%%%%%%%%%%%% BODY OF PAPER %%%%%%%%%%%%%%%%%%
\section{Introduction}\label{sec:intro}
Massive stars (M$\geq$8 M$_{\odot}$) drive powerful stellar feedback that profoundly affects the evolution and star formation efficiency (SFE) of the hosting galaxies. Of such mechanisms, feedback driven by Supernova explosions (SNe) is among the most energetic \citep{2011IAUS..270..247B} and long-lasting \citep[e.g.,][]{leitherer1999, agertz2013}. As the remnant expands, the hot plasma pushes and compresses outwards the atomic and molecular gas in contact with the remnant \citep{chevalier1974} and injects energy, mass and momentum into the interstellar medium (ISM), profoundly affecting its physical properties at multiple spatial scales \citep[see ][for a review]{slane2016}. Mass, energy and momentum are transferred to the nearby material during the adiabatic phase, also known as Sedov-Taylor phase \citep{taylor1950, sedov1959}, during which the energy dissipation is due to expansion and radiative losses are negligible \citep{chevalier1974, cioffi1988, blondin1998, kimOstriker2015, martizzi2015}.\\ At galactic scales, Supernova remnants (SNRs) drive mass-loaded winds that can displace the molecular material, delaying its conversion into stars and hence suppressing star formation in galaxies \citep{bigiel2008, bigiel2010, krumholz2012, leroy2013}. This is known as \textit{negative feedback} \citep{kruijssen2019, kortgen2016}. At the same time, the shock compression of surrounding molecular gas by expanding SNRs can locally (spatial scales $<$10 pc) enhance the density of the molecular material, increase the gas turbulence and eventually trigger the formation of new stars \citep{inutsuka2015,klessen2016}. This effect is known as \textit{positive feedback}. The interplay and relative dominance between positive and negative feedback may depend on several conditions, e.g., the density and gas distribution of the processed material, the evolutionary stage of SNRs \citep{shima2017} and it is paramount in regulating the star formation efficiency and time evolution of galaxies \citep{bigiel2008, bigiel2010, scannapieco2008, leroy2013, heckman2017}. Indeed, it is essential to include stellar feedback in numerical simulations of galactic disc evolution to predict star formation rate and stellar masses comparable to those measured in the ISM \citep{hennebelle2014, smith2018, marinacci2019}.\\ Over time, galaxy evolution simulations have adopted different ad hoc approaches to include SNRs feedback. Early low spatial resolution models treated SN feedback by manually injecting energy into the system at once \citep[for an overview, see][]{ceverino2009}. Such an approach did not consider the Sedov-Taylor phase and as a result, all the injected energy was quickly radiated away with no effects on the ISM \citep{katz1992}. In order to overcome this problem and force the adiabatic phase to occur, later works introduced an artificial delay in the radiative cooling, either by redistributing the injected energy both in space and time \citep{dallavecchia2012} or by switching it off for a certain length of time \citep{stinson2006, governato2010, agertz2011, teyssier2013}. Alternatively to these "delaying cooling" methods, other works treat SN feedback as mechanical feedback and introduce the SNRs at a certain time, with a certain radius and by turning on their kinetic energy and momentum in an ad hoc manner \citep{duboisteyssier2008, kimmcen2014,martizzi2016}. Recent high-resolution simulations within the Feedback In Realistic Environments (FIRE) project are in the process of implementing self-consistently the treatment of SN feedback in galaxy evolution models \citep[e.g., ][]{wetzel2016,sanderson2018}. These works resolve in space and time the different SN evolutionary stages and the different structures of the ISM, limiting the use of sub-resolution approximations for feedback processes \citep{hopkins2014}. Finally, extensive theoretical studies focused on the impact of SN-driven feedback onto the dense molecular material of the ISM have been reported by e.g. \cite{padoan2016,padoan2017} and \cite{seifried2020}.\\ In light of all these extensive theoretical works, current models are able to efficiently describe the expansion of SNRs in a single and/or multi-phase ISM and to make predictions on the energy and momentum imprinted on the nearby material \citep{koo2020}. However, such predictions are still to be fully validated from an observational point of view. In particular, it remains to be constrained the amount of momentum and energy injected by SNRs into the molecular phase of the ISM i.e., the material that primarily fuels star formation in galaxies. This can be efficiently done by studying the emission of those molecular species that trace the high-density shocked gas and whose mm and sub-mm emission is enhanced in sites of SNR-cloud interactions \citep[e.g.,][]{neufeld2007}. Among these species, Silicon Monoxide (SiO) is a unique tracer of dense and shocked molecular material (critical density n$_{crit}>$10$^5$ cm$^{-3}$). Indeed, SiO appears heavily depleted in quiescent regions \citep[$\chi\sim$10$^{-12}$;][]{martinpintado1992,jimenezserra2005} but its abundances can be enhanced by up to a factor $\sim$10$^6$ in regions where the shock propagation causes the sputtering of dust grains or grain-grain collisions. Here, Si is released into the gas phase and SiO is quickly formed \citep{caselli1997, schilke1997, jimenezserra2008, gusdorf2008a}. SiO emission triggered by SNR shocks has been detected toward W51 \citep{dumas2014} and W28 \citep{vaupre2015}, as part of multi-line studies aimed to infer cosmic ray enhancement in SNRs. More recently, in \cite{cosentino2019}, we have reported a dedicated study of the SiO emission arising from the shock interaction between the SNR W44 and the molecular cloud G034.77-00.55 (thereafter G034). Toward this source, the molecular gas pushed away by the expansion of the SNR is interacting with the pre-existing massive molecular cloud, causing a parsec-scale shock seen with relatively narrow SiO emission ($<$3 km s$^{-1}$). The shock is propagating at a velocity of $\sim$ 23 km s$^{-1}$ and is compressing the gas to densities n(H$_2$)$>$10$^5$ cm$^{-3}$ \citep{cosentino2019}. The momentum injected into the dense shocked gas is estimated to be $\sim$20 M$_{\odot}$ km s$^{-1}$ \citep{cosentino2019}.\\
In order to extend the literature sample of SNR-cloud interaction sites seen in SiO emission, we have initiated the ESO-ARO Public Spectroscopic Survey "SHock interactions between supernova REmnants and molecular Clouds" i.e., SHREC, an ongoing large (800 hours) observing program using the 12m antenna at the Arizona Radio Observatory (ARO). SHREC aims to identify sites of ongoing SNR-cloud interaction by mapping the SiO(2-1), H$^{13}$CO$^{+}$(1-0) and HN$^{13}$C(1-0) emission toward a sample of 27 SNRs. These sources have been selected for being relatively nearby (kinematic distance $\leq$ 6 kpc) and for showing evidence of interaction with the surrounding molecular material \citep{ferrand2012,green2019}. This includes enhanced X-ray emission, the presence of OH maser emission at 1720 MHz and enhanced CO(2-1)/CO(1-0) ratios \citep{slane2016}. The final goal of SHREC is to identify sites of large-scale interactions driven by SNRs and to investigate how these affect the star formation potential and dispersal of the surrounding molecular material. The technical presentation of the project and first data release will be presented in a forthcoming paper (Cosentino et al. in prep).\\

\subsection{The SNR IC443}

\begin{figure}
    \centering
    \includegraphics[width=0.365\textwidth,angle=-90,trim=0cm 1cm 0cm 1cm ,clip=True]{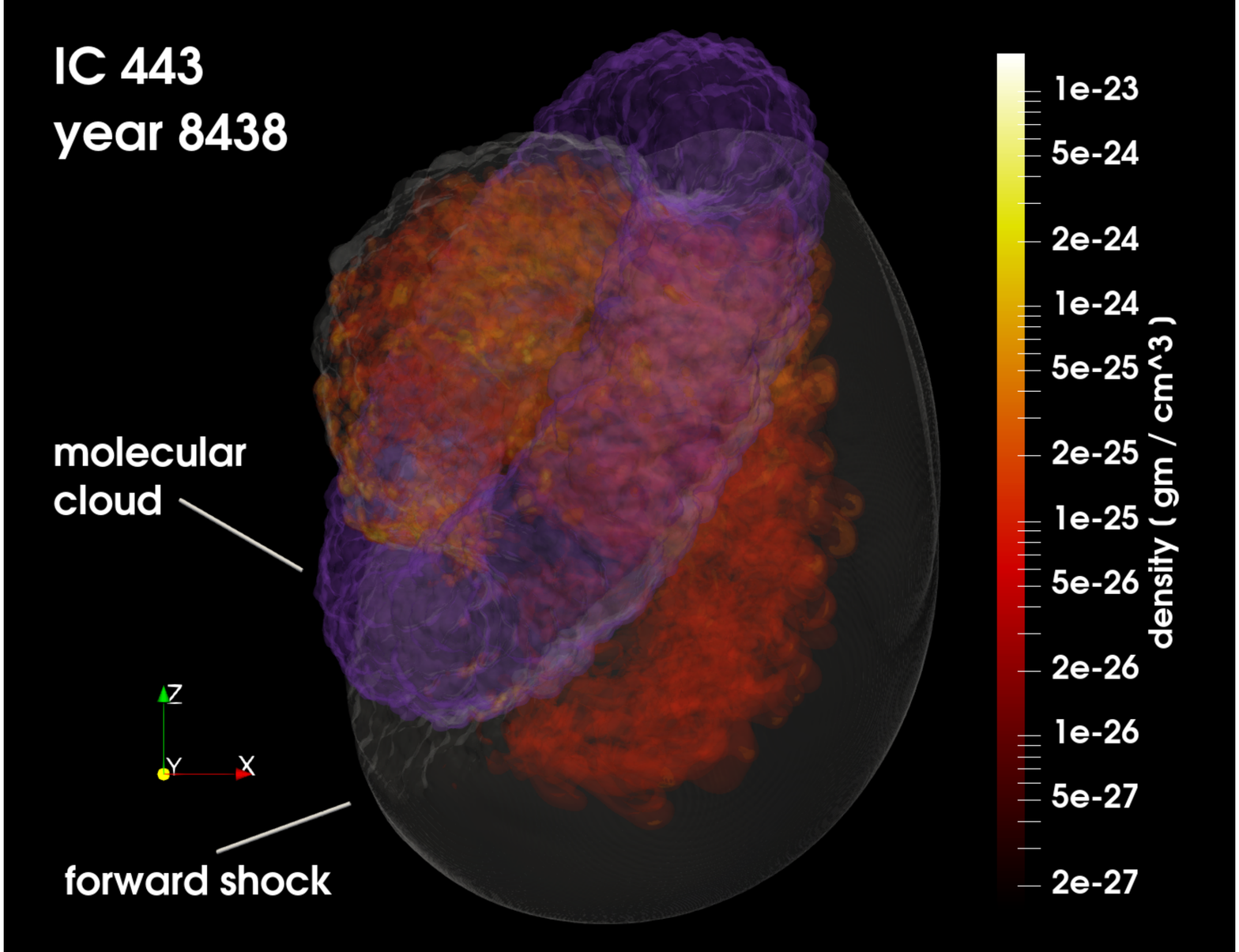}
    \caption{Three-dimensional structure of IC443 derived with a hydrodynamic model that reproduces the morphology of the remnant and the distribution of ejecta \citep{ustamujic2021}. The volume rendering that uses the red colour palette (colour coding on the right of the panel) shows the distribution of ejecta approximately 8400 years after the supernova explosion. The semi-transparent grey outer surface marks the position of the forward shock. The semi-transparent toroidal structure in purple represents the molecular cloud with which the blast wave of the remnant is interacting. The Earth vantage point lies on the negative y-axis i.e., the perspective is in the plane of the sky.
}
    \label{strct3d}
\end{figure}

As part of SHREC, we have obtained SiO(2-1) and H$^{13}$CO$^{+}$(1-0) emission maps toward the well-known source G189.1$+$3.0, also known as IC443. IC443 is a mixed-morphology SNR, i.e., with a shell-like morphology in the radio wavelengths and centrally filled in the X-rays \citep{rho1998}. The source is located at a distance of $\sim$1.9 kpc \citep{ambrociocruz2017} and its age estimate is highly uncertain. Although a typical age of $\sim$30000 years is usually assumed \citep{chevalier1999}, recent simulations suggest that the SNR could be much younger i.e., $\sim$3000-8000 years \citep{troja2008,ustamujic2021}.\\ The IC443 shell is known to be expanding into an atomic cloud in the north-east \citep{denoyer1979} and into a molecular cloud in the north-west to south-east direction \citep{cornett1977}. The first map of the Giant Molecular Cloud surrounding the SNR was reported by \cite{cornett1977} by means of CO(1-0) emission. Later on, \cite{denoyer1979} and \cite{huang1986} identified four major sites of interaction between IC443 and the cloud, named clumps A, B, C and G. By using XMM-Newton maps of the X-ray emission associated with the SNR, \cite{troja2006} reported a geometry of the SNR-cloud system consistent with that of a toroidal molecular cloud wrapped around the expanding SNR and tilted by $\sim$50$^{\circ}$ with respect to the equatorial mid-plane. The three-dimensional structure of the cloud-SNR system is reported in Figure~\ref{strct3d}, as derived with the hydrodynamic model presented by \cite{ustamujic2021}. In the geometry presented in Figure~\ref{strct3d}, clump G corresponds to the part of the torus that is located in the foreground, between the observer and the expanding shell. Among the identified sites, clump G shows the strongest evidence of ongoing shock interaction, i.e., the presence of OH maser emission \citep[e.g. ][]{hewitt2006}, shocked material probed by multi-transitions CO gas \citep[e.g. ][]{zhang2010,dellova2020}, localised non-thermal X-ray emission \citep[e.g. ][]{petre1988,bocchino2000} and shock-excited H$_2$ emission \citep[e.g.,][]{reach2019}. Previous 3mm line survey studies indicated the presence of SiO(2-1) emission toward the clump \citep{ziurys1989,vanDishoeck1993}. In this paper, we present extended maps of the SiO(1-0), SiO(2-1) and H$^{13}$CO$^+$(1-0) emission toward the molecular clump G. We have used the early results of the SHREC large program to study the mass-energy-momentum injection and density enhancement induced by IC443 onto clump G. With this work, we aim to provide a direct estimate of the impact of SN feedback on the molecular phase of the ISM. The paper is organised as follows. In Section~\ref{sec:obs}, we present the observing method and data acquisition. In Section~\ref{sec:results}, we present the result of the analysis performed for the SiO and H$^{13}$CO$^+$ emission toward clump G. Finally, in Sections~\ref{sec:discussion} and \ref{sec:conclusions} we discuss our findings and present our conclusions.

\begin{figure*}
    \centering
    \includegraphics[width=\textwidth]{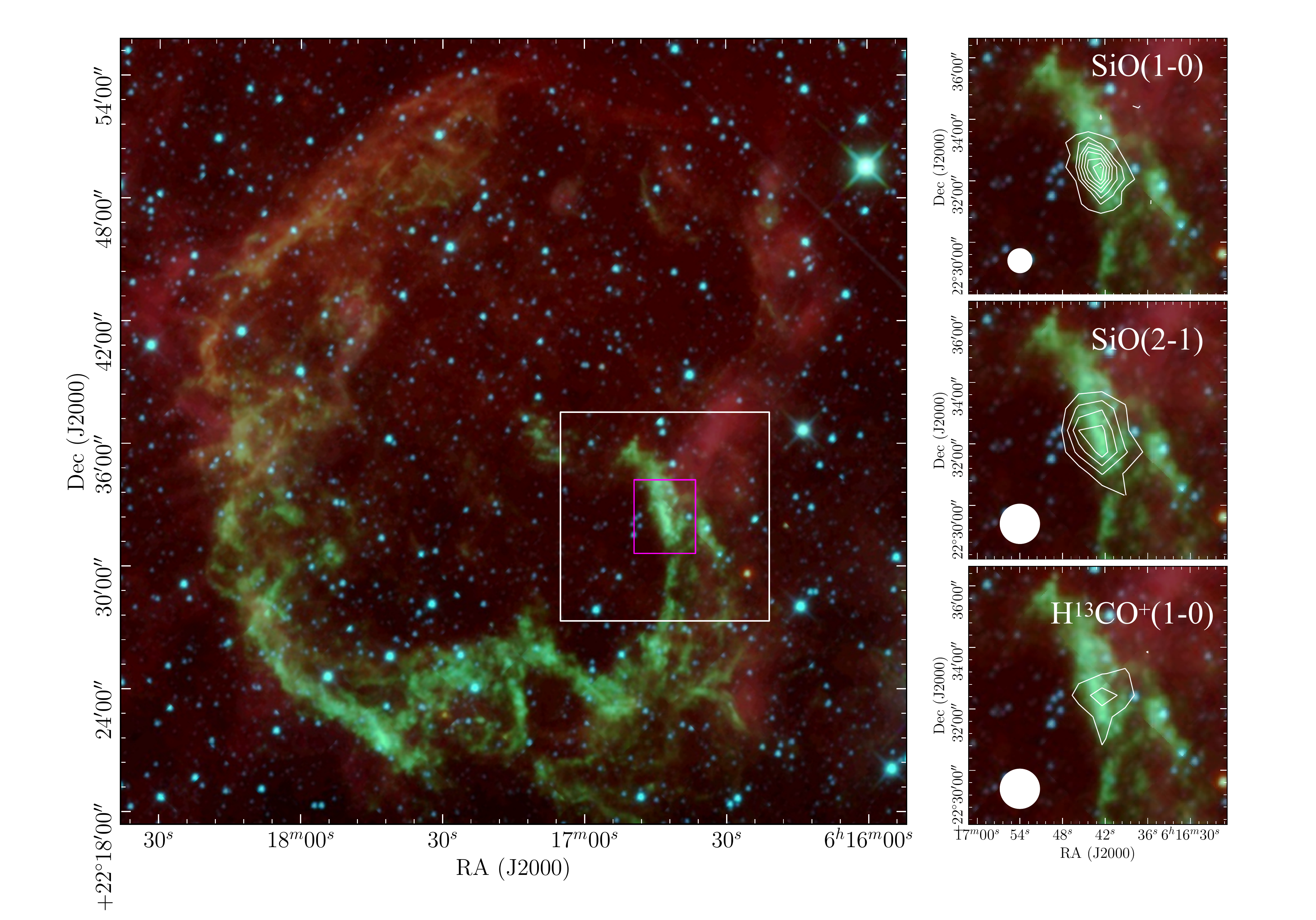}
    \caption{Left: Three-colour image (red= 22 $\mu$m; green = 4.6 $\mu$m; blue = 3.4 $\mu$m ) of the SNR IC443 \citep[WISE all-sky survey][]{wright2010}. White and magenta rectangles indicate the extent of the ARO and Yebes maps respectively. Right: Integrated intensity maps (-22; 6$\,$\kms) of the SiO(1-0) (top; $\sigma$=0.1 K$\,$\kms), SiO(2-1) (middle; $\sigma$=0.2 K$\,$\kms) and H$^{13}$CO$^+$(1-0) (bottom; $\sigma$=0.2 K$\,$\kms) are shown in white contours (from 3$\sigma$ in steps of 3$\sigma$). The ARO and Yebes beam sizes are shown as white circles in the bottom left of each panel.}
    \label{fig1}
\end{figure*}

\section{Observations}\label{sec:obs}
Maps of the SiO(2-1) (86.8469 GHz), H$^{13}$CO$^+$(1-0) (86.7543 GHz) and HN$^{13}$C(1-0) (87.0909 GHz) emission toward clump G were obtained in June 2020 as part of SHREC (P.I. Giuliana Cosentino). Observations were performed using the 12m antenna of the Arizona Radio Observatory (Kitt Peak, Arizona, USA) in on-the-fly (OTF) mode, with scanning speed of 30$^{\prime\prime}$/sec and map size of 10$^{\prime}\times$10$^{\prime}$. The map central coordinates are RA=06$^h$16$^m$32$^s$, Dec=22$^{\circ}$30$^{\prime}$45$^{\prime\prime}$. The AROWS receiver was used with tuning frequency 89.2 GHz and spectral resolution of 78 kHz ($\sim$0.3$\,$\kms \space at 86 GHz), providing a bandwidth of 500 MHz.\\
In November 2020, we used the 40m antenna at the Yebes Observatory (Castilla–La Mancha, Spain) to obtain complementary SiO(1-0) (43.4238 GHz) maps (project code 20B009). The SiO(1-0) observations were performed in OTF mode with scanning speed 15$^{\prime\prime}$/sec, central coordinates RA=06$^h$16$^m$42.4$^s$, Dec=22$^{\circ}$32$^{\prime}$26.3$^{\prime\prime}$ and map size 3.5$^{\prime}\times$3.5$^{\prime}$, corresponding to the full extension of the SiO(2-1) emission. We used the FFT spectrometer in Q band (tuning frequency 43.424 GHz), with spectral resolution of 38 kHz ($\sim$0.3$\,$\kms \space at 43 GHz) and bandwidth 2.5 GHz.\\
For both sets of observations, we use the reference position RA=06$^h$19$^m$01$^s$, Dec=22$^{\circ}$28$^{\prime}$11$^{\prime\prime}$. Intensities were measured in units of antenna temperature and converted into main-beam brightness temperatures using beam efficiencies of 0.61 and 0.52, for the ARO and Yebes observations, respectively. The final data cubes were generated using the GILDAS\footnote{https://www.iram.fr/IRAMFR/GILDAS} package and have beam-sizes of 45$^{\prime\prime}$ and 76$^{\prime\prime}$, for the Yebes and ARO maps, respectively and a common spectral resolution of 0.5$\,$\kms. The achieved root-mean-square (rms) per channel and per beam is of 10 mK for the Yebes maps and and 30 mK for the ARO maps. We note that the HN$^{13}$C(1-0) emission observed as part of SHREC is found to be below the 3$\times$rms level across the full map and hence we do not include it in the following analysis.

\section{Results}\label{sec:results}
\subsection{Morphology and Kinematics of the Shocked Gas}\label{sec:results1}
In Figure~\ref{fig1}, we present the 3-colour image of the SNR IC443 (left) obtained as part of the WISE all-sky survey \citep{wright2010} together with the integrated (in velocity) intensity maps (right) of the SiO(1-0) (top), SiO(2-1) (middle) and H$^{13}$CO$^{+}$(1-0) (bottom) emission. The shocked and dense gas emission is coincident with a bright and extended 4.5 $\mu$m ridge, a signature of shock-excited gas \citep{noriegacrespo2004} and tracing the SNR shock front. The shocked gas tracer emission shows a morphology that is localised with respect to the 4.5 $\mu$m ridge and elongated in the same direction. Such an emission extends >3$\times$ the beam-aperture ($\sim$0.4 and $\sim$0.6 pc for Yebes and ARO observations, respectively) and over parsec scales i.e. $\sim$1.8$\times$1.6 pc$^2$. The H$^{13}$CO$^+$ emission also extends over a parsec scale (1.2$\times$1.3$\,$pc$^2$) but it appears to be more compact than the SiO emission, i.e., <2$\times$ the beam-aperture.

\begin{figure}
    \centering
    \includegraphics[width=0.5\textwidth,trim=0cm 0.5cm 0cm 0cm, clip=true]{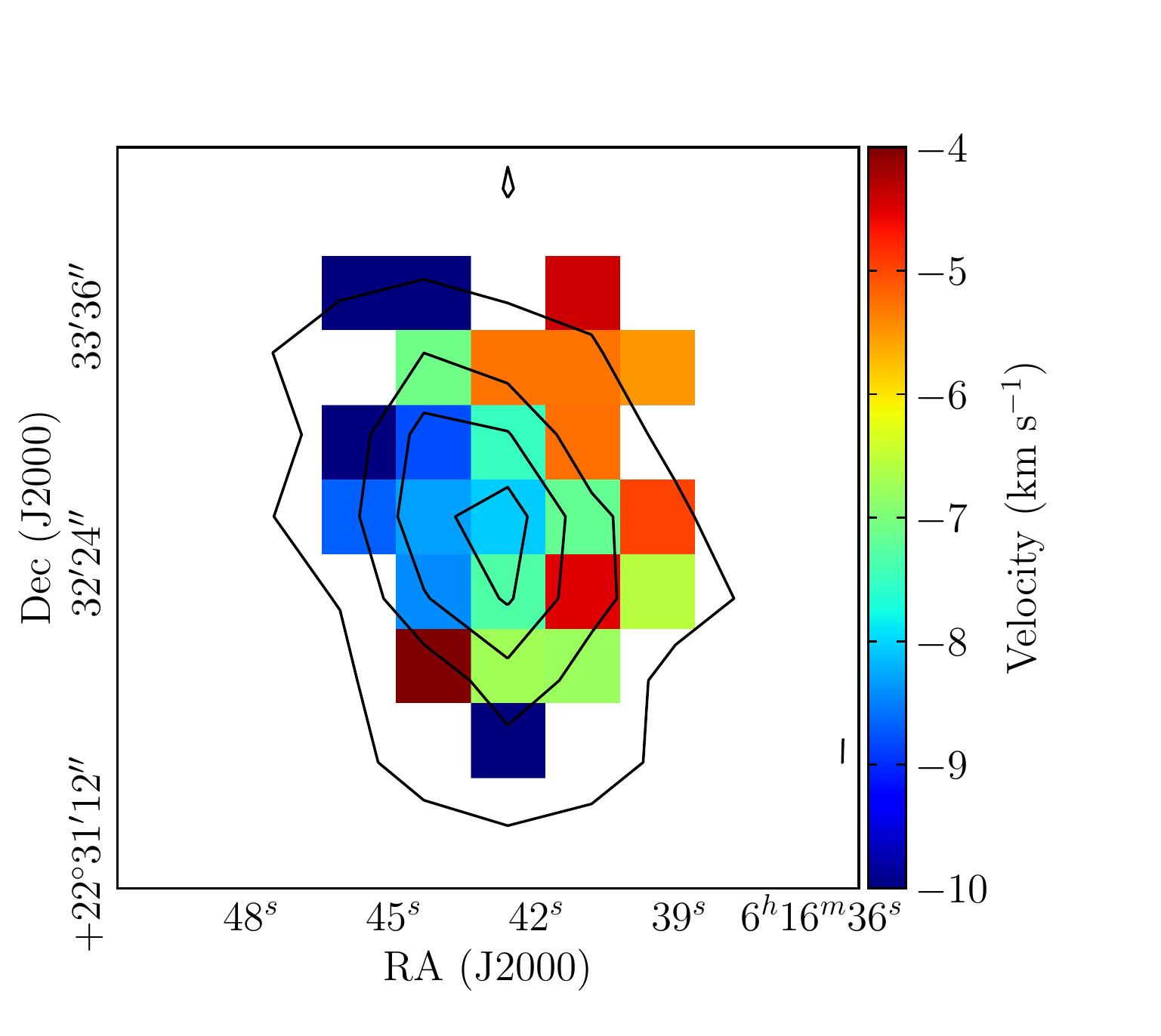}
    \caption{SiO(1-0) moment 1 velocity map (colour scale) toward IC443 superimposed on the SiO(1-0) integrated intensity map contours (black) from 3$\sigma$ by steps of 6$\sigma$ ($\sigma$=0.1 K $\,$\kms).}
    \label{fig2}
\end{figure}

In Figure~\ref{fig2}, we report the moment 1 velocity map (colour scale) obtained for the SiO(1-0), superimposed on the SiO(1-0) integrated intensity emission contours (black). As seen from Figure~\ref{fig2}, the SiO emission shows a velocity gradient, with the blue-shifted gas located toward the east-southeast and the red-shifted emission found toward the west-northwest. Moving away from the 4.5 $\mu$m ridge and into the SNR, the SiO velocity is systematically blue-shifted with respect to the central velocity of IC443 i.e. -4.5 \kms, estimated by means of $^{12}$CO and HCO$^+$ observations \citep{white1987,dickman1992,vanDishoeck1993}.

\subsection{Excitation Conditions of the Shocked Gas}\label{sec:results2}
\begin{figure}
    \centering
    \includegraphics[width=0.5\textwidth,trim=0cm 0.5cm 0cm 0cm, clip=true]{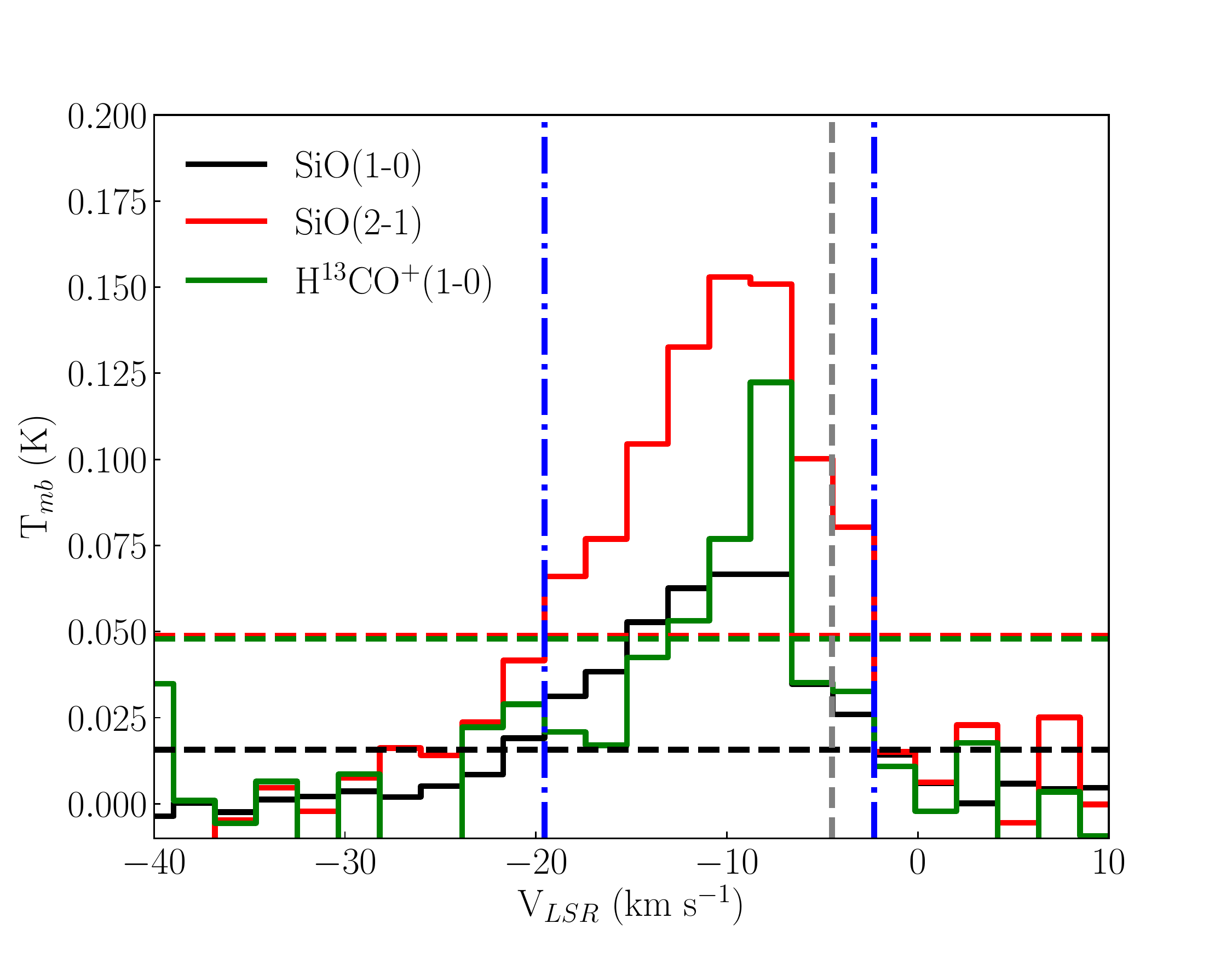}
    \caption{SiO(1-0) (black), SiO(2-1) (red) and H$^{13}$CO$^+$(1-0) (green) spectra extracted toward the regions of the map with emission above the 3$\sigma$ levels. The horizontal dashed lines indicate the 3$\times$rms levels for each species. Vertical blue dashed dotted lines indicated the velocity range considered for the comparison with RADEX models. Finally, the vertical grey line indicates the central velocity of IC443.}
    \label{fig3}
\end{figure}

We now consider the SiO(1-0) and (2-1) line intensities to infer the excitation conditions of the shocked gas, i.e., H$_2$ number density, n(H$_2$), SiO column density, N(SiO), excitation temperature, T$_{ex}$ and how these vary as a function of the gas velocity. For this analysis, the SiO(1-0) emission cube has been spatially smoothed to the same angular resolution of the SiO(2-1) and H$^{13}$CO$^+$(1-0) maps and all cubes were spectrally smoothed to a velocity resolution of 2 km s$^{-1}$. The SiO(1-0) (black), SiO(2-1) (red) and H$^{13}$CO$^+$ (green) spectra obtained by averaging the emission from pixels with signal above 3$\sigma$ are shown in Figure~\ref{fig3}, along with their respective 3$\times$rms levels (dotted horizontal lines).

\begin{figure*}
    \centering
    \includegraphics[width=\textwidth,trim =0cm 2.5cm 0cm 0cm, clip=True]{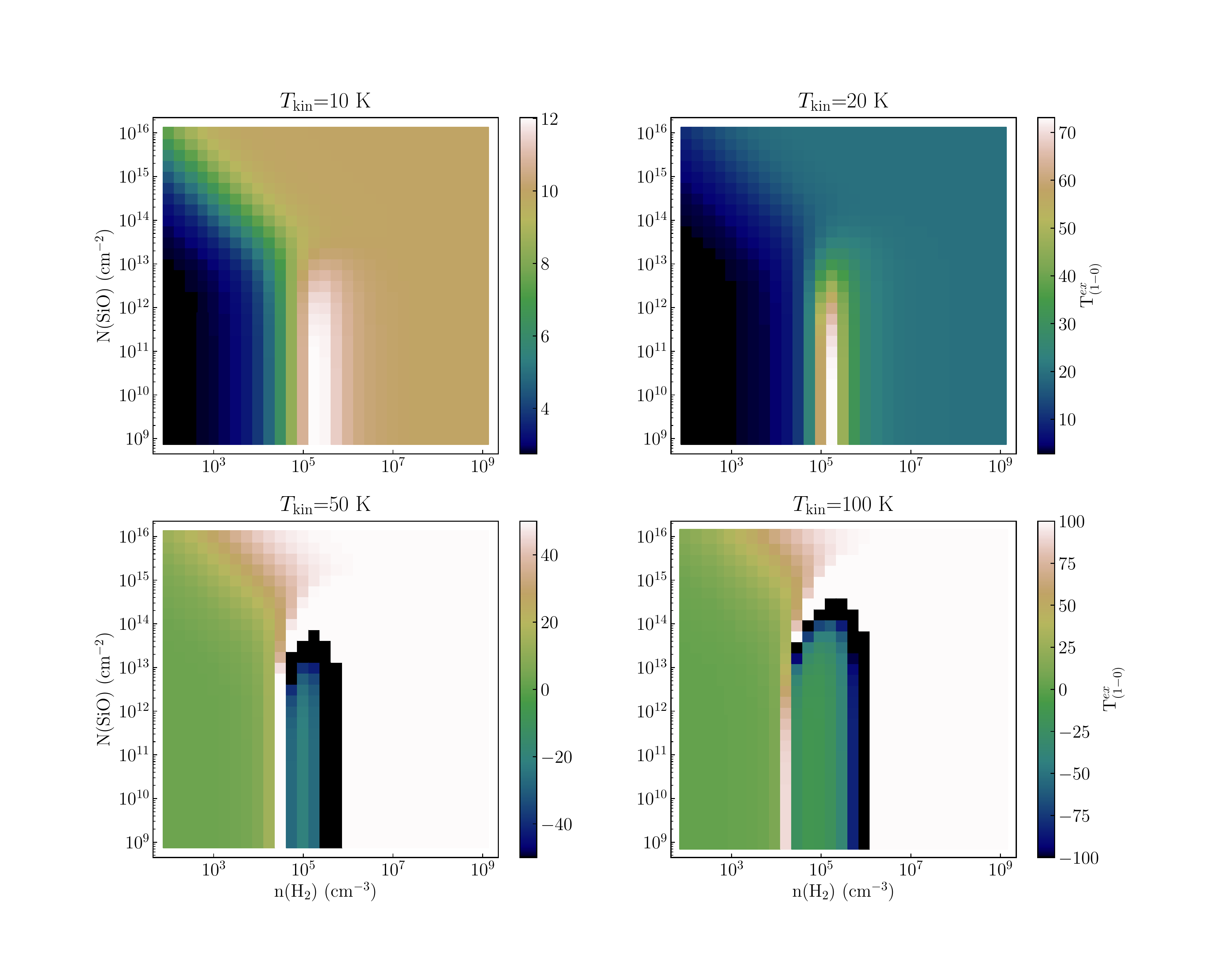}
    \caption{Grids of RADEX models obtained for n(H$_2$) in the range 10$^2$-10$^9$ cm$^{-3}$, N(SiO) in the range 10$^9$-10$^{16}$ cm$^{-2}$ and $T_{\mathrm {kin}}$ of 10 K (top left), 20 K (top right), 50 K (bottom left) and 100 K (bottom right). For each set of N(H$_2$), N(SiO) and $T_{\mathrm {kin}}$, the $T_{\mathrm {ex}}$ predicted for the SiO(1-0) transition is shown in color scales. Although here not shown for simplicity, the $T_{\mathrm {ex}}$ of the SiO(2-1) transitions shows similar trends as a function of kinetic temperature.}
    \label{RADEXtkin}
\end{figure*}

From Figure~\ref{fig3}, the SiO(1-0) and SiO(2-1) emission show significant intensities ($>$3$\times$rms) for velocity channels in the range -19.5,-2.5 km s$^{-1}$ (vertical blue dotted dashed lines). We have therefore limited our analysis to these velocities. For each of the considered velocity channels, we have measured the SiO(1-0) and SiO(2-1) line intensity in unit of K (Table~\ref{tab1}) and used the non-LTE radiative transfer code RADEX \citep{tak2007} to estimate the physical conditions that best reproduce the line strength. RADEX uses the Large Velocity Gradient (LVG) approximation \citep{sobolev1957} to predict line intensities of specific molecules in homogeneous interstellar clouds, starting from a pre-defined system geometry and five input parameters. These are the gas kinetic temperature, $T_{\mathrm {kin}}$, the temperature of the background material, $T_{\mathrm {bg}}$, the volume density of the collisional partners, the molecule column density, N(SiO) and the {width of emission line.} The RADEX output provides the user with line strengths at several frequencies for the selected molecule as well as excitation temperatures and optical depth estimates for each transition. For our analysis, we have assumed a geometry consistent with that of a slab of material processed by a shock. In addition, we have used H$_2$ as collisional partner and specified the H$_2$ volume density, n(H$_2$), as input parameter. The collisional coefficients between H$_2$ and SiO were extracted from the LAMDA database\footnote{https://home.strw.leidenuniv.nl/~moldata/SiO.html} for the first 30 SiO rotational levels \citep{balanca2018}. We have assumed background temperature $T_{\mathrm {bg}}$=2.73 K consistent with the Cosmic Microwave Background emission and used a width of the line of 2 km s$^{-1}$, corresponding to the velocity width of each channel. 
Since only two SiO rotational transitions are here observed, it is not possible to constrain at the same time the three remaining parameters, n(H$_2$), N(SiO) and $T_{\mathrm {kin}}$. Therefore, in the following analysis we proceed by assuming a certain value of $T_{\mathrm {kin}}$ and investigate the sensitivity of our results with respect this assumption. In Figure~\ref{RADEXtkin}, we show multiple grids of RADEX models obtained for n(H$_2$) in the range 10$^2$-10$^9$ cm$^{-3}$, N(SiO) in the range 10$^9$-10$^{16}$ cm$^{-2}$ and with fixed $T_{\mathrm {kin}}$ of 10 K, 20 K, 50 K and 100 K. 

\begin{figure*}
    \centering
    \includegraphics[width=\textwidth,trim=1cm 0cm 1cm 1cm, clip=True]{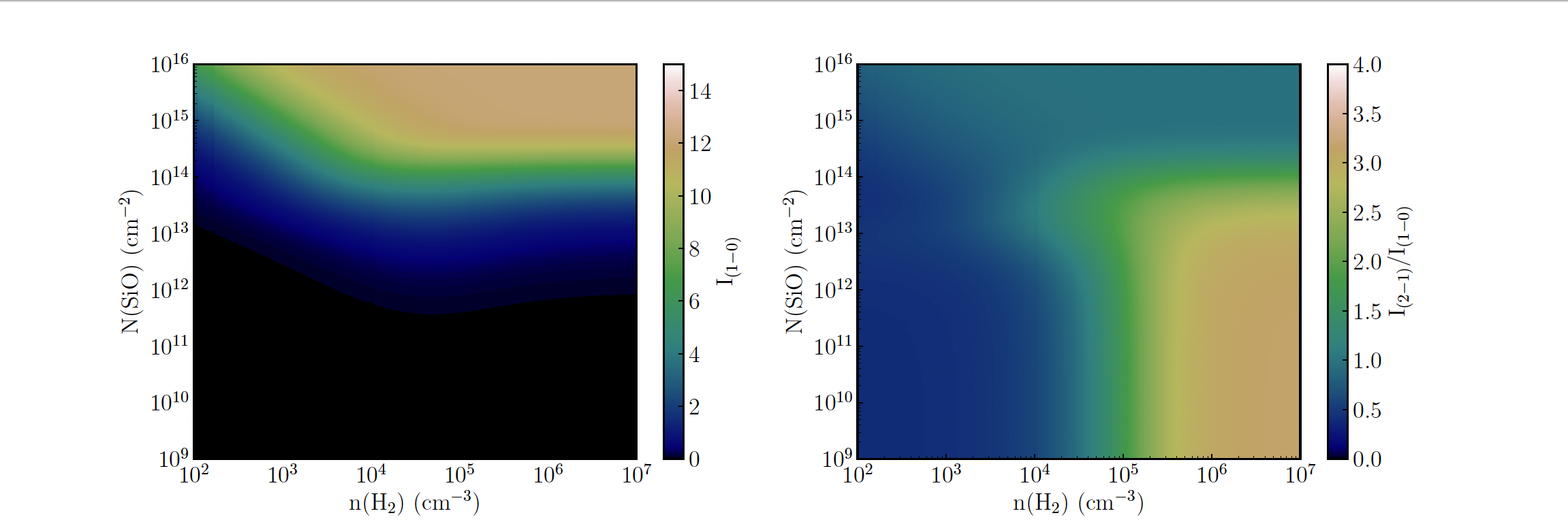}
    \caption{SiO(1-0) line intensities, I$_{(1-0)}$ (left), and SiO(2-1)/SiO(1-0) line intensity ratios, I$_{(2-1)}$/I$_{(1-0)}$ (right) as predicted by the final grid of model obtained for $T_{\mathrm {kin}}$=15 K.}
    \label{RADEX_predictions}
\end{figure*}

As shown in Figure~\ref{RADEXtkin}, the excitation temperature of the SiO(1-0) transition varies significantly for different values of kinetic temperature. In particular, already at $T_{\mathrm {kin}}$>20 K, RADEX predicts negative excitation temperatures for the SiO(1-0) emission i.e., the SiO(1-0) emission is predicted to behave as a maser. SiO(1-0) maser emission is commonly detected toward different objects e.g., variable stars \citep{cho2007}, massive young stellar objects \citep{issaoun2017}, massive star clusters \citep{verheyen2012}, but the rotational transition is usually observed to also be vibrationally excited. At the best of our knowledge, the vibrational ground state of SiO J=1-0 here analysed has been observed with characteristics typical of maser emission only toward evolved stars \citep{boboltz2004}. Toward these objects, the transition shows brightness temperatures $>$10$^3$ K, much higher that the intensities here observed, and is usually detected simultaneously to higher vibrational transitions at close frequencies ($\sim$43.1 GHz). Such frequencies are covered by the bandwidth of our Yebes observations but no vibrationally excited emission are detected. We therefore conclude that the SiO(1-0) emission here reported does not show characteristics typical of maser emission and assume $T_{\mathrm {kin}}$=15 K in the following analysis. Our assumption of $T_{\mathrm {kin}}<$ 20 K reproduces well the observed excitation of the SiO line emission and exclude the possibility that T$_{\mathrm {ex}}$<0 K (maser effects). We also note that the assumed $T_{\mathrm {kin}}$ is consistent with that estimated for the shocked CO emission by \cite{dellova2020} and only a factor of 2 lower than that reported by \cite{ziurys1989}, using multiple Ammonia (NH$_3$) transitions ($T_{\mathrm {kin}}\sim$33 K).\\ Our final grid consists of 250000 models with n(H$_2$) in the range $\sim$10$^2$-10$^7$ cm$^{-3}$ and N(SiO) in the range $\sim$10$^9$-10$^{16}$ cm$^{-2}$, $T_{\mathrm {kin}}$=15 K, $T_{\mathrm {bg}}$=2.73 K and linewidth=2 km s$^{-1}$. The Si(1-0) line intensities, I$_{(1-0)}$ (left panel), and the SiO(2-1)/SiO(1-0) line intensity ratios, I$_{(2-1)}$/I$_{(1-0)}$ (right panel), predicted by our grid of models are shown in Figure~\ref{RADEX_predictions}.

For each of the considered velocity channels, we have compared the measured SiO(1-0) and (2-1) intensities with those predicted by all models in the grid and computed the associated {chi square, $\chi^2$, according} to the following:

\begin{equation}
\chi^2 = \Bigg[\,\frac{(I_{1-0}-I_{RAD})^2}{(\Delta I_{1-0})^2} + \frac{(I_{2-1}-I_{RAD})^2}{(\Delta I_{2-1})^2}\Bigg]\, ,
\label{chisq}
\end{equation}

\noindent
where, $I_{1-0}$ and $I_{2-1}$ are the observed intensities of the SiO(1-0) and (2-1) lines and the subscripts $RAD$ indicates the corresponding quantities estimated by RADEX. For each pair of intensity values, the uncertainty is estimated as the rms per channel, i.e., 5 mK and 15 mK for SiO(1-0) and SiO(2-1), respectively.

\begin{figure*}
    \centering
    \includegraphics[width=\textwidth]{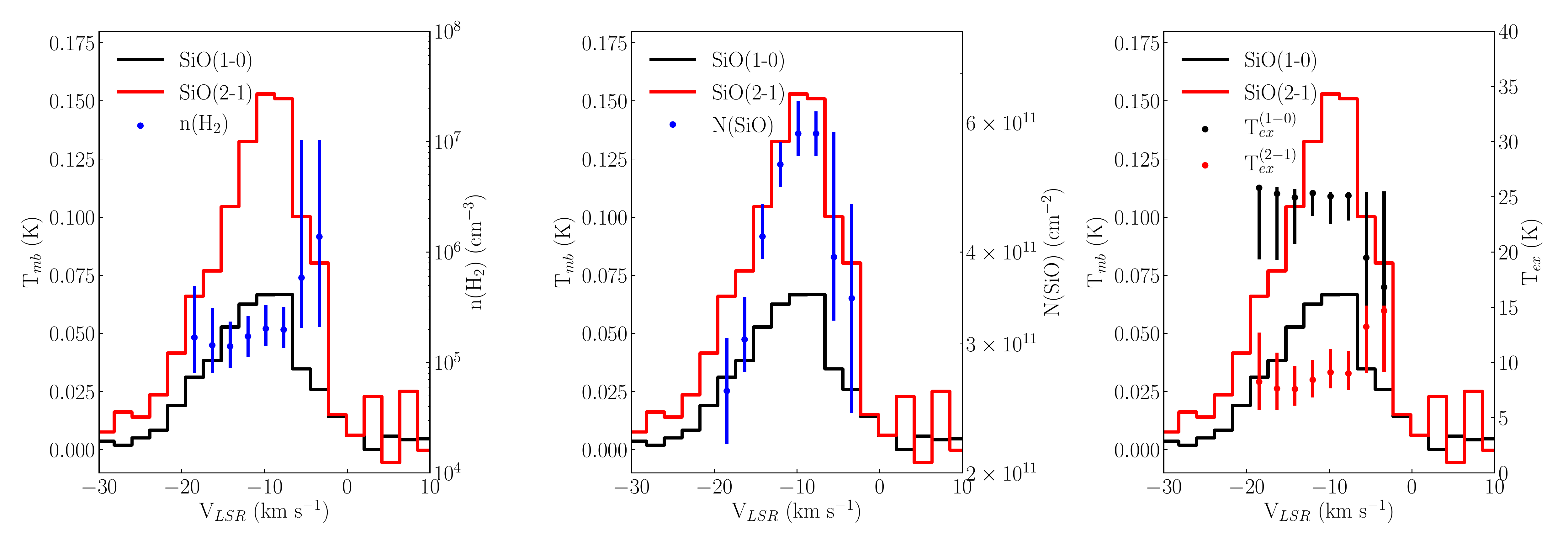}
    \caption{SiO(1-0) (black) and SiO(2-1) (red) spectra as shown in Figure~\ref{fig1} and with superimposed the best values (dot markers) and ranges (vertical lines) obtained for n(H$_2$) (blue, left panel), N(SiO) (blue, middle panel) and the two excitation temperatures T$_{ex}^{1-0}$ (black, right panel) and T$_{ex}^{2-1}$ (red, right panel).} 
    \label{fig4}
\end{figure*}

For each velocity channel, we have extracted a best model as the one that minimises the $\chi^2$ and a range of best RADEX models as those for which $\chi^2$<1. The obtained best  values and ranges are reported in Table~\ref{tab1}, along with the central velocity of the channel and the measured SiO intensities. Since only one transition has been observed for H$^{13}$CO$^+$, a similar analysis is not possible for this dense gas tracer. Hence, we have assumed H$^{13}$CO$^+$(1-0) to have excitation temperatures similar to that estimated for SiO(2-1) and used the best RADEX values at each velocity step to estimate the H$^{13}$CO$^+$ column density. This is justified by the fact that the SiO(2-1) and H$^{13}$CO$^+$(1-0) transitions have a similar critical density. The obtained values are reported in column 9 of Table~\ref{tab1}.\\  
In Figure~\ref{fig4}, the best values (dot markers) and acceptable ranges (vertical lines) obtained for the n(H$_2$) (right panel), N(SiO) (middle panel) and T$_{ex}$ (left panel) are shown along with the SiO(1-0) (black) and SiO(2-1) (red) spectra. From Figure~\ref{fig4}, the n(H$_2$) decreases from red- to blue-shifted velocities. For V$_{LSR}>$-5 km s$^{-1}$, the volume density is $>$5$\times$10$^5$ cm$^{-3}$, while for V$<$-5 km s$^{-1}$, the volume density decreases and sets on a relatively constant value ($\sim$1.5$\times$10$^5$ cm$^{-3}$). A similar trend is observed for the SiO excitation temperatures (right panel), where the two distributions hint to a higher excitation of the gas at red-shifted velocities, i.e., where the  T$_{\mathrm {ex}}^{2-1}$ increases and the T$_{\mathrm {ex}}^{1-0}$ decreases. %We note that the results obtained for the H$_2$ volume density and SiO column density are not dramatically affected by the assumption of $T_{\mathrm {kin}}$=15 K, since similar best values and best ranges are obtained for these parameters even when $T_{\mathrm {kin}}$ up to 100 K is assumed.\\ 

\begin{table*}
    \centering
    \caption{Results of the RADEX analysis performed for the SiO and H$^{13}$CO$^+$ emission. For each velocity channel, the central velocity, the SiO(1-0), SiO(2-1) and H$^{13}$CO$^+$(1-0) intensities, the best n(H$_2$), N(SiO), and T$_{ex}$ values and ranges are reported. We note that all the measured H$^{13}$CO$^+$ intensities are at least higher than 1$\times$ the corresponding rms of 15 mK.}
    \begin{tabular}{cccccccccc}
    \hline
    \hline
        V & I$_{1-0}$  & I$_{2-1}$ & n(H$_2$) & N(SiO) & T$^{1-0}_{ex}$ & T$^{2-1}_{ex}$ & I$_{H^{13}CO^+}$ & N(H$^{13}$CO$^+$)\\
         ($\,$\kms) & (K) & (K) & ($\times$10$^5$ cm$^{-3}$) & ($\times$10$^{11}$ cm$^{-2}$) & (K) & (K) & (K) & ($\times$10$^{10}$ cm$^{-2}$) \\
         \hline
    -18.5 &0.031 &0.066 &1.7  (0.8-4.8)  &2.6 (2.2-3.0) &26 (20-26) &8  (6-13) &0.020 &3.5\\
    -16.5 &0.038 &0.077 &1.4  (0.8-3.0)  &3.0 (2.8-3.5) &25 (19-26) &8  (6-11) &0.017 &3.0\\
    -14.5 &0.053 &0.104 &1.4  (0.9-2.3)  &4.2 (3.9-4.6) &25 (21-26) &8  (6-10) &0.043 &7.0\\
    -12.5 &0.063 &0.133 &1.7  (1.2-2.6)  &5.3 (4.9-5.6) &25 (23-25) &8  (7-10) &0.053 &9.5\\
    -9.5  &0.067 &0.153 &2.0  (1.5-3.2)  &5.8 (5.4-6.4) &25 (23-25) &9  (8-11) &0.077 &13\\
    -7.5  &0.067 &0.151 &2.0  (1.4-3.1)  &5.8 (5.4-6.2) &25 (23-25) &9  (8-11) &0.122 &22\\
    -5.5  &0.035 &0.100 &5.9  (2.1-100)  &3.9 (3.2-5.8) &20 (15-25) &13 (9-15) &0.035 &6.5\\
    -3.5  &0.026 &0.080 &13.8 (2.2-100)  &3.5 (2.4-4.6) &17 (15-25) &15 (9-15) &0.033 &6.3\\
    \hline
    \end{tabular}
    \label{tab1}
    
\end{table*}

From the right panel in Figure~\ref{fig4}, the T$_{ex}$ estimated for the SiO(1-0) transition is $>$15 K at all velocities, i.e, higher than the kinetic temperatures assumed in our models. This \textit{supra-thermal} excitation is a known behaviour, typically observed for linear molecules and at densities consistent with the critical density (n$_{crit}\sim$4$\times$10$^4$ cm$^{-3}$ for the J=1-0 SiO transition). Depending on the H$_2$ and SiO number densities, the radiative and collisional excitation of the SiO compete and LTE conditions are expected to be achieved when collisional excitation dominates. Hence, the T$_{ex}$ is expected to be at most equal to $T_{\mathrm {kin}}$. However, due to quantum selection rules, radiative transitions in linear molecules only occur between successive rotational levels ($\Delta J$=$\pm 1$). Furthermore, the Einstein coefficient A, describing spontaneous radiative decay, increases with the J of the transition. As a consequence of these two effects, higher J-levels will be de-excited faster than the low J-levels, causing a supra-population of the low-J states, and resulting in T$_{ex}$ rising above the LTE value. Such behaviour is analysed in details by \cite{koeppen1980} for the CO molecule.

\subsection{Energy, Mass and Momentum}\label{sec:results3}
As last step in our analysis, we have estimated the mass (M), momentum (P) and kinetic energy (E) of the SiO and H$^{13}$CO$^+$ emission, using the method described in \cite{dierickx2015}:

\begin{equation}
M = \frac{d^2}{\chi(mol)} \times \mu_g m(H_2) \times \Sigma_{pix} N(mol)_{pix},
\end{equation}

\begin{equation}
P = M V,
\end{equation}

\begin{equation}
E = \frac{1}{2} M V^2,
\end{equation}

\noindent
where d is the source kinematic distance \citep[1.9 kpc;][]{ambrociocruz2017}, $\mu_g=$1.36 is the gas molecular weight, m(H$_2$) is the molecular hydrogen mass, $\chi$(mol) is the fractional abundance with respect to H$_2$ and $\Sigma_{pix}$N(mol)$_{pix}$ is the total column density of the molecule, summed for all pixels with signal above 3$\sigma$. Finally, $V$ is the line width at the base (3$\times$rms) of the emission. Similarly to what described in \cite{cosentino2018}, we have obtained an estimate of the $\chi$(SiO) by considering the following equation:

\begin{equation}
\chi(SiO) =  \frac{N(SiO)}{N(H^{13}CO^+)} \times \chi(HCO^+) \times \frac{^{13}C}{^{12}C},
\end{equation}

\noindent
where N(SiO)$\sim$(3.3$\pm$0.5)$\times$10$^{12}$ cm$^{-2}$ and N(H$^{13}$CO$^+$)$\sim$(7 $\pm$1)$\times$10$^{11}$ cm$^{-2}$ have been obtained by summing the values reported in Table~\ref{tab1}. The corresponding uncertainties have been obtained by considering the N(SiO) variability ranges in Table~\ref{tab1} and by assuming that both the SiO and H$^{13}$CO$^+$ column densities have the same relative error. Hence, we have assumed $^{12}$C/$^{13}$C$\sim$50$\pm$10 i.e., we have assigned a 20\% uncertainty \citep[e.g.][]{zeng2017}. We note that the assumed $^{12}$C/$^{13}$C value is consistent with that reported by \cite{dellova2020}, toward clump G. Finally, we assume $\chi$(HCO$^+$)$\sim$(1$\pm$0.5)$\times$10$^{-8}$, as reported by \cite{vanDishoeck1993}. We thus estimated the SiO and H$^{13}$CO$^+$ abundances with respect to H$_2$ to be $\chi$(SiO)$\sim$(1$\pm$0.5)$\times$10$^{-9}$ and $\chi$(H$^{13}$CO$^+$)$\sim$(2$\pm$1)$\times$10$^{-10}$.\\

In order to estimate the SiO and H$^{13}$CO$^+$ column densities at each pixel, we have assumed T$_{ex}\sim$10 K for both species, obtained as the average of the values reported in Table~\ref{tab1} for SiO(2-1). Hence, we use the H$^{13}$CO$^+$(1-0) and SiO(2-1) transitions to estimate the energy, mass and momentum of the dense and shocked gas, respectively. As reported in Table~\ref{tab2}, we obtain M$\sim$125$\pm$50 and 100$\pm$60 M$_{\odot}$, P$\sim$(8$\pm$3)$\times$10$^2$ and P$\sim$(2$\pm$1)$\times$10$^3$ M$_{\odot}$ $\,$\kms and E$\sim$(4.5$\pm$1.8)$\times$10$^{40}$ and $\sim$(2.6$\pm$1.6)$\times$10$^{41}$ ergs for H$^{13}$CO$^+$ and SiO, respectively. The uncertainties have been obtained by following the standard propagation rules. %The SiO M, P and E are a factor of 2-3 lower than the corresponding H$^{13}$CO$^+$ quantities, consistent with the fact that SiO only probes the region where the most powerful shock interaction that sputters dust grains is occurring \citep{caselli1997,schilke1997,jimenezserra2008,gusdorf2008a}.

\begin{table}
    \centering
    \caption{Mass (M), momentum (P) and energy (E) obtained for the shocked and dense gas and corresponding velocity ranges used for the calculation.}
    \begin{tabular}{lllll}
    \hline
    \hline
         Molecule &M &P &E & $\Delta$V\\
                  &(M$_{\odot}$) & (M$_{\odot}$ $\,$\kms) & (erg) & ($\,$\kms) \\
    \hline
         H$^{13}$CO$^+$ &125 $\pm$ 50 &(8 $\pm$ 3) $\times$10$^2$ &(4.5 $\pm$ 1.8) $\times$10$^{40}$ &6\\
         SiO &100 $\pm$ 60 &(2 $\pm$ 1) $\times$10$^3$ &(2.6 $\pm$ 1.6) $\times$10$^{41}$ &16\\
    \hline
    \end{tabular}
    
    \label{tab2}
\end{table}

\section{Discussion} \label{sec:discussion}
\subsection{Can the SiO emission be due to molecular outflows powered by embedded protostars?}
The shock interaction between IC443 and clump G has been largely investigated from both an observational \citep[e.g.][]{ziurys1989,dickman1992,vanDishoeck1993,reach2019,dellova2020,kokusho2020} and theoretical \citep[e.g.][]{troja2006,troja2008,ustamujic2021} point of view. However, SiO is usually widely observed in regions of ongoing star formation activity \citep[e.g., ][]{codella2007,lopezsepulcre2011,liu2020}. Hence, we now discuss the possibility that the SiO emission observed toward clump G may be due to molecular outflows driven by deeply embedded protostars.\\
As reported in Table~\ref{tab2}, the energy measured for the shocked gas is $\sim$2.6$\times$10$^{41}$ ergs. This is several orders of magnitudes lower than the typical kinetic energy measured for molecular outflows powered by high-mass protostars i.e., $\sim$10$^{46}$ ergs \citep[e.g., ][]{zhang2005,lopezsepulcre2009} and intermediate-mass protostars i.e., $\sim$10$^{43}$-10$^{44}$ ergs \citep[e.g., ][]{beltran2006,beltran2008}. We therefore exclude the possibility that ongoing high-mass and intermediate-mass star formation may be driving the observed SiO emission. On the other hand, \cite{dellova2020} reported the presence of $\sim$25 Young Stellar Objects (YSOs) spatially associated with the shocked gas in clump G and located within a distance of $\pm$500 pc \citep[see Figure 14 in][]{dellova2020}. We now assume that the SiO emission is entirely due to putative outflows powered by these 25 YSOs, and that each source contributes equally to the final SiO emission. In this scenario, each outflow should have on average a mass of M$_{SiO}$/25$\sim$4 M$_{\odot}$ and a momentum of P$_{SiO}$/25$\sim$80 M$_{\odot}$ km s$^{-1}$. These estimates are several orders of magnitude higher than those typically measured toward molecular outflows driven by low-mass protostars i.e.,  M$\sim$0.005$-$0.15 M$_{\odot}$, P$\sim$0.004$-$0.12M$_{\odot}$ km s$^{-1}$ \citep[e.g.,][]{arce2014}. 
This is even more stringent if we consider the more likely scenario in which only few of the 25 YSOs are effectively driving outflows and thus contributing to the observed SiO emission. Hence, although we cannot exclude that a small contribution to the observed SiO emission may be due to molecular outflows powered by low-mass protostars, the major contribution to the observed SiO emission likely arises from the large-scale shock interaction occurring between IC443 and clump G.

\subsection{SiO as probe of the shock interaction between IC443 and clump G: positive feedback driven by SNRs}\label{positive}
The presence of such a large-scale shock interaction is further supported by the kinematic structure observed for the SiO emission and reported in Section~\ref{sec:results1}. The SiO emission is indeed significantly blue-shifted with respect to the central velocity of the clump \citep{dickman1992} and it presents a global velocity gradient with the blue-shifted emission appearing toward the east-southeast and the red-shifted gas located toward the west-northwest.
Such a kinematic structure cannot be reproduced by a collection of molecular outflows driven by low-mass protostars, which have been seen to be randomly oriented with respect to the parental clump, in star forming regions \citep{dunham2016,stephens2017}. On the contrary, the well-organised SiO kinematic structure is similar to that reported by \cite{cosentino2019} toward the molecular cloud G034 known to be interacting with the SNR W44 \citep{wootten1977}.
Toward this region, the SiO emission is seen to be blue-shifted with respect to the central velocity of the cloud \citep[42 km s$^{-1}$][]{cosentino2019} and spatially associated with a 4.5 $\mu$m extended ridge \citep{cosentino2018}.
The SiO emission toward G034 as seen with ALMA, shows a sharp gradient of 2-3 \kms, within 3$^{\prime\prime}$, which is followed by a shallower gradient of 5-6 \kms, across $>$10$^{\prime\prime}$ scales (equivalent to linear scales of 0.15 pc). Our ARO and Yebes observations probe spatial scales of 0.7-0.4 pc and therefore cannot resolve with such detail the observed SiO emission. However, IC443 is 1 kpc closer than W44 ($\sim$2.9 kpc) and its SiO emission is almost a factor of 2 more extended, which allows us to appreciate a clear shocked gas velocity gradient across the ridge. The highly blue-shifted SiO gas here observed is naturally explained when the geometry suggested by \cite{troja2006} is considered. In this scenario, clump G is located in the foreground with respect to IC443 and hence the shock wave released by the SNR hits the clump from behind pushing and dragging the shocked gas toward the observer. The fact that the SiO blue-shifted emission appears directed toward the inner part of IC443, is likely due to the fact that the shock is impacting on the cloud with a certain angle with respect to the line of sight. This was first suggested by \cite{dickman1992} and \cite{vanDishoeck1993} and is consistent with what was reported by \cite{reach2019}. These authors modelled the interaction between IC443 and clump G as occurring through two CJ-type shocks of $\sim$60 $\,$\kms and $\sim$37 $\,$\kms \space and dynamical age 5$\times$10$^3$ and 3$\times$10$^3$ years, respectively and that are hitting the cloud with angles of $\sim$60-65$^{\circ}$ with respect to the line of sight.\\ The multiple shocks driven by IC443 and impacting on clump G may be responsible for the H$_2$ volume density profile reported in Figure~\ref{fig4}. The higher densities seen at velocity $\geq$-5 km s$^{-1}$ may be associated with the initial stronger impact between the shocks and the cloud, from which the bulk of the SiO emission is likely arising. After this first compression, the gas is dragged and decelerates toward the observer, appearing as highly blue-shifted. This is also supported by the fact that the higher excitation of the shocked gas also occurs at velocity $\geq$-5 km s$^{-1}$ and decreases at more blue-shifted velocities.\\ 
From Figure~\ref{fig4}, the H$_2$ volume density of the shocked gas toward clump G is $\geq$10$^5$ cm$^{-3}$ at all velocities, consistent with both the SiO(1-0) and (2-1) critical densities. These values are comparable to those required to ignite star formation in the ISM \citep[e.g.,][]{parmentier2011}.\\ Toward clump G, several studies have reported H$_2$ volume densities of the pre-shocked gas in the range n(H$_2$)$\sim$10$^3$-10$^4$ cm$^{-3}$ \citep{vanDishoeck1993,dellova2020} i.e., slightly lower than those typically observed in dark clouds. This suggests that the shock propagation enhances the gas density by more than a factor of 10 and up to a factor of 100. By using XMM-Newton observations of clump G, \cite{troja2006} identified a strong X-ray absorption and reported a n(H) column density variation, along the line of sight, of 5 $\times$ 10$^{21}$ cm$^{-2}$. By considering such a variation and the post-shocked H$_2$ gas density here measured, i.e., 10$^5$ cm$^{-3}$, the length of the shocked region can be estimated as $\sim$2.5$\times$10$^{16}$ cm (or 0.008 pc). When a shock velocity of 25 km s$^{-1}$ is considered \citep{dickman1992}, the time since the first shock interaction is therefore of $\sim$300 years. For such a time-scale, a factor of 10 density enhancement in the post-shocked gas can be explained as due to both the shock propagation, the presence of radiative cooling processes and significant energy dissipation by particles acceleration. The presence of such mechanisms is indicated by the detection of non-thermal X-ray emission, toward clump G \citep[e.g.,][]{bocchino2000}. A factor of 100 is instead well beyond the typical density enhancements caused by shock propagation. We therefore suggest that the pre-shocked gas density toward clump G is at least of n(H$_2$)$\sim$10$^4$ cm$^{-3}$. This supports the idea that clump G may have have been a coherent dense structure pre-existent to the SNe event.\\
Finally, we note that the high-density measured in the post-shocked gas may help to explain the enhanced $\gamma$-ray emission measured toward clump G \citep{albert2007,acciari2009,abdo2010}. 

\subsection{The H$^{13}$CO$^+$(1-0) emission toward clump G: shock chemistry product or molecular cloud in the making?}
Emission from H$^{13}$CO$^+$ is a good probe of the dense gas distribution in molecular clouds \citep[e.g.,][]{vasyunina2011}. As shown in Figure~\ref{fig1}, the H$^{13}$CO$^+$ emission is spatially coincident with the shocked gas emission but less extended. No significant H$^{13}$CO$^+$ emission is detected outside the shocked region, supporting the low-density values measured for the ambient gas \citep{vanDishoeck1993}. In addition, the H$^{13}$CO$^+$ spectrum reported in Figure~\ref{fig3} shows a profile similar to that of both the SiO(1-0) and (2-1) transitions. The spatial and spectral similarities between the SiO and H$^{13}$CO$^+$ emission hint toward a common nature of the two species. In this scenario, the H$^{13}$CO$^+$ emission is likely a consequence of the ongoing shock chemistry. Indeed, emission from ions such as HCO$^+$ is known to be enhanced either in the earliest stages of the shock \citep[see Figure 5 in][]{flower2003} or in the far post-shock gas, when the temperatures have gone down to $\sim$30 K \citep{bergin1996}. This scenario is consistent with the idea that the cloud was pre-existent with respect to the SNR, as already suggested by \cite{dickman1992}, more recently discussed by \cite{ustamujic2021} and as discussed in Section~\ref{positive}. This is also supported by the fact that the dense and shocked gas mass estimates are similar and that the emission is spatially localised with respect to the 4.5 $\mu$m ridge, direct probe of the shock front.
\\ Alternatively to this scenario, the H$^{13}$CO$^+$ emission may be probing the material of clump G that is being shock-compressed with a process similar to that described by \cite{inutsuka2015}. Here, bubbles due to stellar feedback expand into the clumpy multi-phase ISM, driving multiple episodes of shock compression into the nearby low-density material participating in the assembling of dark clouds. In the IC443 scenario, pre-existent low-density material of clump G may have been be compressed by the propagating shocks to densities sufficient to enable the collisional excitation of the H$^{13}$CO$^+$(1-0). This may be supported by the fact that the dense gas mass reported in Table~\ref{tab2} is comparable to the mass of the ambient gas measured from multiple CO transitions by \cite{dellova2020}. However, we note that the peak velocity of the H$^{13}$CO$^+$ emission ($\sim$-8 km s$^{-1}$) does not coincide with that of the ambient cloud in \cite{dellova2020} ($\sim$-3.5 km s$^{-1}$). We therefore suggest that the H$^{13}$CO$^+$ emission is mainly due to the ongoing shock chemistry in IC443. We note that the dense gas mass here estimated from H$^{13}$CO$^+$ is a factor of 2-3 higher than that reported by \cite{dickman1992} for clump G, i.e., $\sim$ 40 M$_{\odot}$, obtained from HCO$^+$ emission. However, these authors did not take into account possible optical depth effects in the HCO$^+$ emission.

\subsection{Negative feedback driven by IC443: comparing observations with model predictions.}
In Section~\ref{sec:results3}, we have investigated the mass, energy and momentum calculated for both the dense and shocked gas detected toward clump G. As reported in Table~\ref{tab2}, we estimate a mass of the shocked gas of $\sim$100$\pm$60 M$_{\odot}$. Considering the length of the shocked region reported in Section~\ref{positive}, i.e., 2.5$\times$10$^{16}$ cm or 0.008 pc, the volume of the shocked region is 1.8$\times$1.6$\times$0.008 pc$^3$. For the measured post-shocked gas density of n(H$_2$)$\sim$10$^5$ cm$^{-3}$, the mass enclosed in such a volume is $\sim$90 M$_{\odot}$. This provide an independent confirmation to the values here reported.\\
IC443 is known to be interacting with the molecular material toward three additional sites, named clumps A, B, C \citep{dickman1992}. Assuming that the momentum transferred into clumps A, B and C is equal to that measured toward clump G, we estimate a momentum transferred from the SNR into the surrounding \textit{molecular} material in the range $\sim$3.2-8$\times$10$^3$ M$_{\odot}$ \kms. Since the strongest interaction is known to be occurring toward clump G \citep{claussen1997}, such a value should be regarded as an upper limit.\\
State-of-the-art numerical simulations predict the amount of momentum transferred from an expanding SNR into the nearby ISM to be (1-5)$\times$10$^5$ M$_{\odot}$ \kms, when a kinetic energy of 10$^{51}$ ergs released by the SNR is assumed \citep{kimOstriker2015,li2015,martizzi2015,iffrighennebelle2015,zhangchevalier2019}. Assuming such kinetic energy for IC443 \citep{ustamujic2021}, we estimate that the momentum carried away by the interaction between the SNR shocks and the surrounding molecular material represents $<$10\% of the total imprinted momentum.\\
More in general, considering the momentum of the shocked and dense gas reported in Table~\ref{tab2}, 2$\times$10$^3$ M$_{\odot}$ \kms, and the SiO and H$^{13}$CO$^+$ emission spatial coverage of 1.8$\times$1.6 pc$^2$ and 1.3$\times$1.2 pc$^2$ respectively, we estimate the \textit{momentum per unit area} of the gas to be $\sim$500-690 M$_{\odot}$ \kms pc$^2$. For this estimate, we assume the gas emission to be plane parallel and along the line of sight. We note that the momentum per unit area here estimated should also be regarded as an upper limit, since the interaction toward other sites is likely to be weaker than that observed toward clump G. Considering for the SNR a diameter of 45$^{\prime\prime}$ \citep{green2019}, we estimate the area of the bubble to be $\sim$1800 pc$^2$. Since the molecular material surrounding IC443 is distributed as a toroid around the expanding bubble \citep{troja2006}, we assume that the SNR effective area that is directly in contact with molecular material is $\sim$20\% and in any case $<$50\%, since IC443 is expanding into an atomic cloud in the North. Therefore, we estimate the momentum carried by the molecular material to be $\sim$1.8-2.5$\times$10$^5$ M$_{\odot}$ \kms. This is 35$-$50\% the momentum typically injected by a SNR.\\ Our calculation indicates that the molecular material can be a relevant carrier of the momentum injected by SNR feedback into the ISM. The importance of this resides in the fact that the cold dense molecular material of the ISM is the primary fuel of star formation in galaxies. The imprinted momentum contributes to maintain the level of turbulence in the ISM \citep{padoan2016}, a key ingredient in the star formation process. Finally, we note that the momentum deposited by SNRs into the ISM is further increased by the presence of accelerated comic rays. Since their energy is not radiated away during the Sedov-Taylor phase, cosmic ray further support the SNR expansion and prolong the momentum deposition phase. As a result, the momentum injected by SNRs into the nearby material can be boosted by a factor of 5-10, for density of the ISM $>$10$^2$ cm$^{-3}$. For a more detailed discussion we refer to \cite{Diesing2018}.

\section{Conclusions}\label{sec:conclusions}
In this paper, we have used SiO(2-1) and H$^{13}$CO$^+$(1-0) observations obtained as part of SHREC as well as complementary SiO(1-0) observations obtained by 40m antenna at the Yebes Observatory to investigate the negative and positive feedback driven by the SNR IC443 onto the molecular clump G. Our results can be summarised as follows:
\begin{itemize}
    \item[i)] The SiO emission shows an elongated morphology, spatially coincident and parallel to an extended ridge of shocked gas seen at 4.5 $\mu$m. The SiO kinematics is organised as a well-ordered structure, with the shocked material being systematically blue shifted with respect to the central velocity of the SNR.\\
    \item[ii)] The shocked gas kinematic structure as well as its inferred mass (100 M$_{\odot}$), momentum (2$\times$10$^3$ M$_{\odot}$ \kms) and  energy (2.6$\times$10$^{41}$ ergs) cannot be solely explained as the product of ongoing star formation activity in clump G. Therefore we conclude that the bulk of the SiO emission arises from the ongoing shock interaction between the clump and IC443.\\
    \item[iii)] Toward clump G, the shock propagation enhances the gas density to values n(H$_2$)$\geq$10$^5$ cm$^{-3}$, a factor >10 higher than the density of the pre-shocked material and consistent with the densities required to ignite star formation in molecular clouds.\\
    \item[iv)] The dense gas mass estimated from the H$^{13}$CO$^+$ emission is similar to that estimated for the shocked gas. Furthermore, the dense gas emission is spatially concentrated toward the 4.5 $\mu$m ridge. We interpret this result as evidence that the H$^{13}$CO$^+$ emission is likely due to shock chemistry effects and that clump G was pre-existent with respect to IC443.\\
    \item[v)] Finally, we estimate that between 35-50\% of the momentum injected by IC443 is transferred to the molecular phase of the ISM, making the molecular material an important momentum carrier in sites of SNR-cloud interactions. The injected momentum helps maintain turbulence in the molecular ISM that fuels star formation in galaxies.
\end{itemize}

\section*{Acknowledgements}
We thank Prof. Floris van der Tak for providing us with helpful insights on the matter of the SiO supra-thermal excitation. G.C. and P.G. acknowledges support from a Chalmers Cosmic Origins postdoctoral fellowship. I.J.-S. has received partial support from the Spanish State Research Agency (AEI; project number PID2019-105552RB-C41). R.F. acknowledges funding from the European Union’s Horizon 2020 research and innovation programme under the Marie Sklodowska-Curie grant agreement No 101032092. J.C.T. acknowledges support from ERC project 788829– MSTAR. S.V. acknowledges partial funding from the European Research Council (ERC) Advanced Grant MOPPEX 833460. S.Z. acknowledges support from NAOJ ALMA Scientific Research Grant Number 2016-03B.

%%%%%%%%%%%%%%%%%%%%%%%%%%%%%%%%%%%%%%%%%%%%%%%%%%
\section*{Data Availability}
The SiO and H$^{13}$CO$^+$ data cubes used for the study here presented will be distributed upon request to the corresponding author. Data obtained as part of SHREC will also be made publicly available through dedicated website in the coming months.



\bibliographystyle{mnras}
\bibliography{example} % if your bibtex file is called example.bib

\begin{thebibliography}{}
\makeatletter
\relax
\def\mn@urlcharsother{\let\do\@makeother \do\$\do\&\do\#\do\^\do\_\do\%\do\~}
\def\mn@doi{\begingroup\mn@urlcharsother \@ifnextchar [ {\mn@doi@}
  {\mn@doi@[]}}
\def\mn@doi@[#1]#2{\def\@tempa{#1}\ifx\@tempa\@empty \href
  {http://dx.doi.org/#2} {doi:#2}\else \href {http://dx.doi.org/#2} {#1}\fi
  \endgroup}
\def\mn@eprint#1#2{\mn@eprint@#1:#2::\@nil}
\def\mn@eprint@arXiv#1{\href {http://arxiv.org/abs/#1} {{\tt arXiv:#1}}}
\def\mn@eprint@dblp#1{\href {http://dblp.uni-trier.de/rec/bibtex/#1.xml}
  {dblp:#1}}
\def\mn@eprint@#1:#2:#3:#4\@nil{\def\@tempa {#1}\def\@tempb {#2}\def\@tempc
  {#3}\ifx \@tempc \@empty \let \@tempc \@tempb \let \@tempb \@tempa \fi \ifx
  \@tempb \@empty \def\@tempb {arXiv}\fi \@ifundefined
  {mn@eprint@\@tempb}{\@tempb:\@tempc}{\expandafter \expandafter \csname
  mn@eprint@\@tempb\endcsname \expandafter{\@tempc}}}

\bibitem[\protect\citeauthoryear{{Abdo} et~al.,}{{Abdo}
  et~al.}{2010}]{abdo2010}
{Abdo} A.~A.,  et~al., 2010, \mn@doi [\apj] {10.1088/0004-637X/712/1/459},
  \href {https://ui.adsabs.harvard.edu/abs/2010ApJ...712..459A} {712, 459}

\bibitem[\protect\citeauthoryear{{Acciari} et~al.,}{{Acciari}
  et~al.}{2009}]{acciari2009}
{Acciari} V.~A.,  et~al., 2009, \mn@doi [\apjl] {10.1088/0004-637X/698/2/L133},
  \href {https://ui.adsabs.harvard.edu/abs/2009ApJ...698L.133A} {698, L133}

\bibitem[\protect\citeauthoryear{{Agertz}, {Teyssier}  \& {Moore}}{{Agertz}
  et~al.}{2011}]{agertz2011}
{Agertz} O.,  {Teyssier} R.,   {Moore} B.,  2011, \mn@doi [\mnras]
  {10.1111/j.1365-2966.2010.17530.x}, \href
  {https://ui.adsabs.harvard.edu/abs/2011MNRAS.410.1391A} {410, 1391}

\bibitem[\protect\citeauthoryear{{Agertz}, {Kravtsov}, {Leitner}  \&
  {Gnedin}}{{Agertz} et~al.}{2013}]{agertz2013}
{Agertz} O.,  {Kravtsov} A.~V.,  {Leitner} S.~N.,   {Gnedin} N.~Y.,  2013,
  \mn@doi [\apj] {10.1088/0004-637X/770/1/25}, \href
  {https://ui.adsabs.harvard.edu/abs/2013ApJ...770...25A} {770, 25}

\bibitem[\protect\citeauthoryear{{Albert} et~al.,}{{Albert}
  et~al.}{2007}]{albert2007}
{Albert} J.,  et~al., 2007, \mn@doi [\apjl] {10.1086/520957}, \href
  {https://ui.adsabs.harvard.edu/abs/2007ApJ...664L..87A} {664, L87}

\bibitem[\protect\citeauthoryear{{Ambrocio-Cruz}, {Rosado}, {de la Fuente},
  {Silva}  \& {Blanco-Pi{\~n}on}}{{Ambrocio-Cruz}
  et~al.}{2017}]{ambrociocruz2017}
{Ambrocio-Cruz} P.,  {Rosado} M.,  {de la Fuente} E.,  {Silva} R.,
  {Blanco-Pi{\~n}on} A.,  2017, \mn@doi [\mnras] {10.1093/mnras/stx1936}, \href
  {https://ui.adsabs.harvard.edu/abs/2017MNRAS.472...51A} {472, 51}

\bibitem[\protect\citeauthoryear{{Balan{\c{c}}a}, {Dayou}, {Faure},
  {Wiesenfeld}  \& {Feautrier}}{{Balan{\c{c}}a} et~al.}{2018}]{balanca2018}
{Balan{\c{c}}a} C.,  {Dayou} F.,  {Faure} A.,  {Wiesenfeld} L.,   {Feautrier}
  N.,  2018, \mn@doi [\mnras] {10.1093/mnras/sty1681}, \href
  {https://ui.adsabs.harvard.edu/abs/2018MNRAS.479.2692B} {479, 2692}

\bibitem[\protect\citeauthoryear{{Bally}}{{Bally}}{2011}]{2011IAUS..270..247B}
{Bally} J.,  2011, in {Alves} J.,  {Elmegreen} B.~G.,  {Girart} J.~M.,
  {Trimble} V.,  eds, ~ Vol. 270, Computational Star Formation. pp 247--254,
  \mn@doi{10.1017/S1743921311000469}

\bibitem[\protect\citeauthoryear{{Beltr{\'a}n}, {Girart}  \&
  {Estalella}}{{Beltr{\'a}n} et~al.}{2006}]{beltran2006}
{Beltr{\'a}n} M.~T.,  {Girart} J.~M.,   {Estalella} R.,  2006, \mn@doi [\aap]
  {10.1051/0004-6361:20065132}, \href
  {https://ui.adsabs.harvard.edu/abs/2006A&A...457..865B} {457, 865}

\bibitem[\protect\citeauthoryear{{Beltr{\'a}n}, {Estalella}, {Girart}, {Ho}  \&
  {Anglada}}{{Beltr{\'a}n} et~al.}{2008}]{beltran2008}
{Beltr{\'a}n} M.~T.,  {Estalella} R.,  {Girart} J.~M.,  {Ho} P.~T.~P.,
  {Anglada} G.,  2008, \mn@doi [\aap] {10.1051/0004-6361:20078045}, \href
  {https://ui.adsabs.harvard.edu/abs/2008A&A...481...93B} {481, 93}

\bibitem[\protect\citeauthoryear{{Bergin}, {Snell}  \& {Goldsmith}}{{Bergin}
  et~al.}{1996}]{bergin1996}
{Bergin} E.~A.,  {Snell} R.~L.,   {Goldsmith} P.~F.,  1996, \mn@doi [\apj]
  {10.1086/176974}, \href
  {https://ui.adsabs.harvard.edu/abs/1996ApJ...460..343B} {460, 343}

\bibitem[\protect\citeauthoryear{{Bigiel}, {Leroy}, {Walter}, {Brinks}, {de
  Blok}, {Madore}  \& {Thornley}}{{Bigiel} et~al.}{2008}]{bigiel2008}
{Bigiel} F.,  {Leroy} A.,  {Walter} F.,  {Brinks} E.,  {de Blok} W.~J.~G.,
  {Madore} B.,   {Thornley} M.~D.,  2008, \mn@doi [\aj]
  {10.1088/0004-6256/136/6/2846}, \href
  {https://ui.adsabs.harvard.edu/abs/2008AJ....136.2846B} {136, 2846}

\bibitem[\protect\citeauthoryear{{Bigiel}, {Leroy}, {Walter}, {Blitz},
  {Brinks}, {de Blok}  \& {Madore}}{{Bigiel} et~al.}{2010}]{bigiel2010}
{Bigiel} F.,  {Leroy} A.,  {Walter} F.,  {Blitz} L.,  {Brinks} E.,  {de Blok}
  W.~J.~G.,   {Madore} B.,  2010, \mn@doi [\aj] {10.1088/0004-6256/140/5/1194},
  \href {https://ui.adsabs.harvard.edu/abs/2010AJ....140.1194B} {140, 1194}

\bibitem[\protect\citeauthoryear{{Blondin}, {Wright}, {Borkowski}  \&
  {Reynolds}}{{Blondin} et~al.}{1998}]{blondin1998}
{Blondin} J.~M.,  {Wright} E.~B.,  {Borkowski} K.~J.,   {Reynolds} S.~P.,
  1998, \mn@doi [\apj] {10.1086/305708}, \href
  {https://ui.adsabs.harvard.edu/abs/1998ApJ...500..342B} {500, 342}

\bibitem[\protect\citeauthoryear{{Boboltz} \& {Claussen}}{{Boboltz} \&
  {Claussen}}{2004}]{boboltz2004}
{Boboltz} D.~A.,  {Claussen} M.~J.,  2004, \mn@doi [\apj] {10.1086/386541},
  \href {https://ui.adsabs.harvard.edu/abs/2004ApJ...608..480B} {608, 480}

\bibitem[\protect\citeauthoryear{{Bocchino} \& {Bykov}}{{Bocchino} \&
  {Bykov}}{2000}]{bocchino2000}
{Bocchino} F.,  {Bykov} A.~M.,  2000, \aap, \href
  {https://ui.adsabs.harvard.edu/abs/2000A&A...362L..29B} {362, L29}

\bibitem[\protect\citeauthoryear{{Caselli}, {Hartquist}  \& {Havnes}}{{Caselli}
  et~al.}{1997}]{caselli1997}
{Caselli} P.,  {Hartquist} T.~W.,   {Havnes} O.,  1997, \aap, \href
  {https://ui.adsabs.harvard.edu/abs/1997A&A...322..296C} {322, 296}

\bibitem[\protect\citeauthoryear{{Ceverino} \& {Klypin}}{{Ceverino} \&
  {Klypin}}{2009}]{ceverino2009}
{Ceverino} D.,  {Klypin} A.,  2009, \mn@doi [\apj]
  {10.1088/0004-637X/695/1/292}, \href
  {https://ui.adsabs.harvard.edu/abs/2009ApJ...695..292C} {695, 292}

\bibitem[\protect\citeauthoryear{{Chevalier}}{{Chevalier}}{1974}]{chevalier1974}
{Chevalier} R.~A.,  1974, \mn@doi [\apj] {10.1086/152740}, \href
  {https://ui.adsabs.harvard.edu/abs/1974ApJ...188..501C} {188, 501}

\bibitem[\protect\citeauthoryear{{Chevalier}}{{Chevalier}}{1999}]{chevalier1999}
{Chevalier} R.~A.,  1999, \mn@doi [\apj] {10.1086/306710}, \href
  {https://ui.adsabs.harvard.edu/abs/1999ApJ...511..798C} {511, 798}

\bibitem[\protect\citeauthoryear{Cho, Lee  \& Park}{Cho et~al.}{2007}]{cho2007}
Cho S.-H.,  Lee C.~W.,   Park Y.-S.,  2007, \mn@doi [The Astrophysical Journal]
  {10.1086/510837}, 657, 482

\bibitem[\protect\citeauthoryear{{Cioffi}, {McKee}  \& {Bertschinger}}{{Cioffi}
  et~al.}{1988}]{cioffi1988}
{Cioffi} D.~F.,  {McKee} C.~F.,   {Bertschinger} E.,  1988, \mn@doi [\apj]
  {10.1086/166834}, \href
  {https://ui.adsabs.harvard.edu/abs/1988ApJ...334..252C} {334, 252}

\bibitem[\protect\citeauthoryear{{Claussen}, {Frail}, {Goss}  \&
  {Gaume}}{{Claussen} et~al.}{1997}]{claussen1997}
{Claussen} M.~J.,  {Frail} D.~A.,  {Goss} W.~M.,   {Gaume} R.~A.,  1997,
  \mn@doi [\apj] {10.1086/304784}, \href
  {https://ui.adsabs.harvard.edu/abs/1997ApJ...489..143C} {489, 143}

\bibitem[\protect\citeauthoryear{{Codella}, {Cabrit}, {Gueth}, {Cesaroni},
  {Bacciotti}, {Lefloch}  \& {McCaughrean}}{{Codella}
  et~al.}{2007}]{codella2007}
{Codella} C.,  {Cabrit} S.,  {Gueth} F.,  {Cesaroni} R.,  {Bacciotti} F.,
  {Lefloch} B.,   {McCaughrean} M.~J.,  2007, \mn@doi [\aap]
  {10.1051/0004-6361:20066800}, \href
  {https://ui.adsabs.harvard.edu/abs/2007A&A...462L..53C} {462, L53}

\bibitem[\protect\citeauthoryear{{Cornett}, {Chin}  \& {Knapp}}{{Cornett}
  et~al.}{1977}]{cornett1977}
{Cornett} R.~H.,  {Chin} G.,   {Knapp} G.~R.,  1977, \aap, \href
  {https://ui.adsabs.harvard.edu/abs/1977A&A....54..889C} {54, 889}

\bibitem[\protect\citeauthoryear{{Cosentino} et~al.,}{{Cosentino}
  et~al.}{2018}]{cosentino2018}
{Cosentino} G.,  et~al., 2018, \mn@doi [\mnras] {10.1093/mnras/stx3013}, \href
  {https://ui.adsabs.harvard.edu/abs/2018MNRAS.474.3760C} {474, 3760}

\bibitem[\protect\citeauthoryear{{Cosentino} et~al.,}{{Cosentino}
  et~al.}{2019}]{cosentino2019}
{Cosentino} G.,  et~al., 2019, \mn@doi [\apjl] {10.3847/2041-8213/ab38c5},
  \href {https://ui.adsabs.harvard.edu/abs/2019ApJ...881L..42C} {881, L42}

\bibitem[\protect\citeauthoryear{{Dalla Vecchia} \& {Schaye}}{{Dalla Vecchia}
  \& {Schaye}}{2012}]{dallavecchia2012}
{Dalla Vecchia} C.,  {Schaye} J.,  2012, \mn@doi [\mnras]
  {10.1111/j.1365-2966.2012.21704.x}, \href
  {https://ui.adsabs.harvard.edu/abs/2012MNRAS.426..140D} {426, 140}

\bibitem[\protect\citeauthoryear{{Dell'Ova} et~al.,}{{Dell'Ova}
  et~al.}{2020}]{dellova2020}
{Dell'Ova} P.,  et~al., 2020, \mn@doi [\aap] {10.1051/0004-6361/202038339},
  \href {https://ui.adsabs.harvard.edu/abs/2020A&A...644A..64D} {644, A64}

\bibitem[\protect\citeauthoryear{{Denoyer}}{{Denoyer}}{1979}]{denoyer1979}
{Denoyer} L.~K.,  1979, \mn@doi [\apjl] {10.1086/183057}, \href
  {https://ui.adsabs.harvard.edu/abs/1979ApJ...232L.165D} {232, L165}

\bibitem[\protect\citeauthoryear{{Dickman}, {Snell}, {Ziurys}  \&
  {Huang}}{{Dickman} et~al.}{1992}]{dickman1992}
{Dickman} R.~L.,  {Snell} R.~L.,  {Ziurys} L.~M.,   {Huang} Y.-L.,  1992,
  \mn@doi [\apj] {10.1086/171987}, \href
  {https://ui.adsabs.harvard.edu/abs/1992ApJ...400..203D} {400, 203}

\bibitem[\protect\citeauthoryear{{Dierickx}, {Jim{\'e}nez-Serra}, {Rivilla}  \&
  {Zhang}}{{Dierickx} et~al.}{2015}]{dierickx2015}
{Dierickx} M.,  {Jim{\'e}nez-Serra} I.,  {Rivilla} V.~M.,   {Zhang} Q.,  2015,
  \mn@doi [\apj] {10.1088/0004-637X/803/2/89}, \href
  {https://ui.adsabs.harvard.edu/abs/2015ApJ...803...89D} {803, 89}

\bibitem[\protect\citeauthoryear{{Diesing} \& {Caprioli}}{{Diesing} \&
  {Caprioli}}{2018}]{Diesing2018}
{Diesing} R.,  {Caprioli} D.,  2018, \mn@doi [\prl]
  {10.1103/PhysRevLett.121.091101}, \href
  {https://ui.adsabs.harvard.edu/abs/2018PhRvL.121i1101D} {121, 091101}

\bibitem[\protect\citeauthoryear{{Dubois} \& {Teyssier}}{{Dubois} \&
  {Teyssier}}{2008}]{duboisteyssier2008}
{Dubois} Y.,  {Teyssier} R.,  2008, \mn@doi [\aap]
  {10.1051/0004-6361:20078326}, \href
  {https://ui.adsabs.harvard.edu/abs/2008A&A...477...79D} {477, 79}

\bibitem[\protect\citeauthoryear{{Dumas}, {Vaupr{\'e}}, {Ceccarelli},
  {Hily-Blant}, {Dubus}, {Montmerle}  \& {Gabici}}{{Dumas}
  et~al.}{2014}]{dumas2014}
{Dumas} G.,  {Vaupr{\'e}} S.,  {Ceccarelli} C.,  {Hily-Blant} P.,  {Dubus} G.,
  {Montmerle} T.,   {Gabici} S.,  2014, \mn@doi [\apjl]
  {10.1088/2041-8205/786/2/L24}, \href
  {https://ui.adsabs.harvard.edu/abs/2014ApJ...786L..24D} {786, L24}

\bibitem[\protect\citeauthoryear{{Dunham}, {Arce}, {Mardones}, {Lee},
  {Matthews}, {Stutz}  \& {Williams}}{{Dunham} et~al.}{2014}]{arce2014}
{Dunham} M.~M.,  {Arce} H.~G.,  {Mardones} D.,  {Lee} J.-E.,  {Matthews} B.~C.,
   {Stutz} A.~M.,   {Williams} J.~P.,  2014, \mn@doi [\apj]
  {10.1088/0004-637X/783/1/29}, \href
  {https://ui.adsabs.harvard.edu/abs/2014ApJ...783...29D} {783, 29}

\bibitem[\protect\citeauthoryear{{Dunham} et~al.,}{{Dunham}
  et~al.}{2016}]{dunham2016}
{Dunham} M.~M.,  et~al., 2016, \mn@doi [\apj] {10.3847/0004-637X/823/2/160},
  \href {https://ui.adsabs.harvard.edu/abs/2016ApJ...823..160D} {823, 160}

\bibitem[\protect\citeauthoryear{{Ferrand} \& {Safi-Harb}}{{Ferrand} \&
  {Safi-Harb}}{2012}]{ferrand2012}
{Ferrand} G.,  {Safi-Harb} S.,  2012, \mn@doi [Advances in Space Research]
  {10.1016/j.asr.2012.02.004}, \href
  {https://ui.adsabs.harvard.edu/abs/2012AdSpR..49.1313F} {49, 1313}

\bibitem[\protect\citeauthoryear{{Flower} \& {Pineau des For{\^e}ts}}{{Flower}
  \& {Pineau des For{\^e}ts}}{2003}]{flower2003}
{Flower} D.~R.,  {Pineau des For{\^e}ts} G.,  2003, \mn@doi [\mnras]
  {10.1046/j.1365-8711.2003.06716.x}, \href
  {https://ui.adsabs.harvard.edu/abs/2003MNRAS.343..390F} {343, 390}

\bibitem[\protect\citeauthoryear{{Governato} et~al.,}{{Governato}
  et~al.}{2010}]{governato2010}
{Governato} F.,  et~al., 2010, \mn@doi [\nat] {10.1038/nature08640}, \href
  {https://ui.adsabs.harvard.edu/abs/2010Natur.463..203G} {463, 203}

\bibitem[\protect\citeauthoryear{{Green}}{{Green}}{2019}]{green2019}
{Green} D.~A.,  2019, \mn@doi [Journal of Astrophysics and Astronomy]
  {10.1007/s12036-019-9601-6}, \href
  {https://ui.adsabs.harvard.edu/abs/2019JApA...40...36G} {40, 36}

\bibitem[\protect\citeauthoryear{{Gusdorf}, {Cabrit}, {Flower}  \& {Pineau Des
  For{\^e}ts}}{{Gusdorf} et~al.}{2008}]{gusdorf2008a}
{Gusdorf} A.,  {Cabrit} S.,  {Flower} D.~R.,   {Pineau Des For{\^e}ts} G.,
  2008, \mn@doi [\aap] {10.1051/0004-6361:20078900}, \href
  {https://ui.adsabs.harvard.edu/abs/2008A&A...482..809G} {482, 809}

\bibitem[\protect\citeauthoryear{{Heckman} \& {Thompson}}{{Heckman} \&
  {Thompson}}{2017}]{heckman2017}
{Heckman} T.~M.,  {Thompson} T.~A.,  2017, arXiv e-prints, \href
  {https://ui.adsabs.harvard.edu/abs/2017arXiv170109062H} {p. arXiv:1701.09062}

\bibitem[\protect\citeauthoryear{{Hennebelle} \& {Iffrig}}{{Hennebelle} \&
  {Iffrig}}{2014}]{hennebelle2014}
{Hennebelle} P.,  {Iffrig} O.,  2014, \mn@doi [\aap]
  {10.1051/0004-6361/201423392}, \href
  {https://ui.adsabs.harvard.edu/abs/2014A&A...570A..81H} {570, A81}

\bibitem[\protect\citeauthoryear{{Hewitt}, {Yusef-Zadeh}, {Wardle}, {Roberts}
  \& {Kassim}}{{Hewitt} et~al.}{2006}]{hewitt2006}
{Hewitt} J.~W.,  {Yusef-Zadeh} F.,  {Wardle} M.,  {Roberts} D.~A.,   {Kassim}
  N.~E.,  2006, \mn@doi [\apj] {10.1086/508331}, \href
  {https://ui.adsabs.harvard.edu/abs/2006ApJ...652.1288H} {652, 1288}

\bibitem[\protect\citeauthoryear{{Hopkins}, {Kere{\v{s}}}, {O{\~n}orbe},
  {Faucher-Gigu{\`e}re}, {Quataert}, {Murray}  \& {Bullock}}{{Hopkins}
  et~al.}{2014}]{hopkins2014}
{Hopkins} P.~F.,  {Kere{\v{s}}} D.,  {O{\~n}orbe} J.,  {Faucher-Gigu{\`e}re}
  C.-A.,  {Quataert} E.,  {Murray} N.,   {Bullock} J.~S.,  2014, \mn@doi
  [\mnras] {10.1093/mnras/stu1738}, \href
  {https://ui.adsabs.harvard.edu/abs/2014MNRAS.445..581H} {445, 581}

\bibitem[\protect\citeauthoryear{{Huang} \& {Thaddeus}}{{Huang} \&
  {Thaddeus}}{1986}]{huang1986}
{Huang} Y.~L.,  {Thaddeus} P.,  1986, \mn@doi [\apj] {10.1086/164649}, \href
  {https://ui.adsabs.harvard.edu/abs/1986ApJ...309..804H} {309, 804}

\bibitem[\protect\citeauthoryear{{Iffrig} \& {Hennebelle}}{{Iffrig} \&
  {Hennebelle}}{2015}]{iffrighennebelle2015}
{Iffrig} O.,  {Hennebelle} P.,  2015, \mn@doi [\aap]
  {10.1051/0004-6361/201424556}, \href
  {https://ui.adsabs.harvard.edu/abs/2015A&A...576A..95I} {576, A95}

\bibitem[\protect\citeauthoryear{{Inutsuka}, {Inoue}, {Iwasaki}  \&
  {Hosokawa}}{{Inutsuka} et~al.}{2015}]{inutsuka2015}
{Inutsuka} S.-i.,  {Inoue} T.,  {Iwasaki} K.,   {Hosokawa} T.,  2015, \mn@doi
  [\aap] {10.1051/0004-6361/201425584}, \href
  {https://ui.adsabs.harvard.edu/abs/2015A&A...580A..49I} {580, A49}

\bibitem[\protect\citeauthoryear{{Issaoun} et~al.,}{{Issaoun}
  et~al.}{2017}]{issaoun2017}
{Issaoun} S.,  et~al., 2017, \mn@doi [\aap] {10.1051/0004-6361/201731548},
  \href {https://ui.adsabs.harvard.edu/abs/2017A&A...606A.126I} {606, A126}

\bibitem[\protect\citeauthoryear{{Jim{\'e}nez-Serra}, {Mart{\'\i}n-Pintado},
  {Rodr{\'\i}guez-Franco}  \& {Mart{\'\i}n}}{{Jim{\'e}nez-Serra}
  et~al.}{2005}]{jimenezserra2005}
{Jim{\'e}nez-Serra} I.,  {Mart{\'\i}n-Pintado} J.,  {Rodr{\'\i}guez-Franco} A.,
    {Mart{\'\i}n} S.,  2005, \mn@doi [\apjl] {10.1086/432467}, \href
  {https://ui.adsabs.harvard.edu/abs/2005ApJ...627L.121J} {627, L121}

\bibitem[\protect\citeauthoryear{{Jim{\'e}nez-Serra}, {Caselli},
  {Mart{\'\i}n-Pintado}  \& {Hartquist}}{{Jim{\'e}nez-Serra}
  et~al.}{2008}]{jimenezserra2008}
{Jim{\'e}nez-Serra} I.,  {Caselli} P.,  {Mart{\'\i}n-Pintado} J.,   {Hartquist}
  T.~W.,  2008, \mn@doi [\aap] {10.1051/0004-6361:20078054}, \href
  {https://ui.adsabs.harvard.edu/abs/2008A&A...482..549J} {482, 549}

\bibitem[\protect\citeauthoryear{{Katz}}{{Katz}}{1992}]{katz1992}
{Katz} N.,  1992, \mn@doi [\apj] {10.1086/171366}, \href
  {https://ui.adsabs.harvard.edu/abs/1992ApJ...391..502K} {391, 502}

\bibitem[\protect\citeauthoryear{{Kim} \& {Ostriker}}{{Kim} \&
  {Ostriker}}{2015}]{kimOstriker2015}
{Kim} C.-G.,  {Ostriker} E.~C.,  2015, \mn@doi [\apj]
  {10.1088/0004-637X/815/1/67}, \href
  {https://ui.adsabs.harvard.edu/abs/2015ApJ...815...67K} {815, 67}

\bibitem[\protect\citeauthoryear{{Kimm} \& {Cen}}{{Kimm} \&
  {Cen}}{2014}]{kimmcen2014}
{Kimm} T.,  {Cen} R.,  2014, \mn@doi [\apj] {10.1088/0004-637X/788/2/121},
  \href {https://ui.adsabs.harvard.edu/abs/2014ApJ...788..121K} {788, 121}

\bibitem[\protect\citeauthoryear{{Klessen} \& {Glover}}{{Klessen} \&
  {Glover}}{2016}]{klessen2016}
{Klessen} R.~S.,  {Glover} S. C.~O.,  2016, \mn@doi [Saas-Fee Advanced Course]
  {10.1007/978-3-662-47890-5\_2}, \href
  {https://ui.adsabs.harvard.edu/abs/2016SAAS...43...85K} {43, 85}

\bibitem[\protect\citeauthoryear{{Koeppen} \& {Kegel}}{{Koeppen} \&
  {Kegel}}{1980}]{koeppen1980}
{Koeppen} J.,  {Kegel} W.~H.,  1980, \aaps, \href
  {https://ui.adsabs.harvard.edu/abs/1980A&AS...42...59K} {42, 59}

\bibitem[\protect\citeauthoryear{{Kokusho}, {Torii}, {Nagayama}, {Kaneda},
  {Sano}, {Ishihara}  \& {Onaka}}{{Kokusho} et~al.}{2020}]{kokusho2020}
{Kokusho} T.,  {Torii} H.,  {Nagayama} T.,  {Kaneda} H.,  {Sano} H.,
  {Ishihara} D.,   {Onaka} T.,  2020, \mn@doi [\apj]
  {10.3847/1538-4357/ab9cb3}, \href
  {https://ui.adsabs.harvard.edu/abs/2020ApJ...899...49K} {899, 49}

\bibitem[\protect\citeauthoryear{{Koo}, {Kim}, {Park}  \& {Ostriker}}{{Koo}
  et~al.}{2020}]{koo2020}
{Koo} B.-C.,  {Kim} C.-G.,  {Park} S.,   {Ostriker} E.~C.,  2020, \mn@doi
  [\apj] {10.3847/1538-4357/abc1e7}, \href
  {https://ui.adsabs.harvard.edu/abs/2020ApJ...905...35K} {905, 35}

\bibitem[\protect\citeauthoryear{{K{\"o}rtgen}, {Seifried}, {Banerjee},
  {V{\'a}zquez-Semadeni}  \& {Zamora-Avil{\'e}s}}{{K{\"o}rtgen}
  et~al.}{2016}]{kortgen2016}
{K{\"o}rtgen} B.,  {Seifried} D.,  {Banerjee} R.,  {V{\'a}zquez-Semadeni} E.,
  {Zamora-Avil{\'e}s} M.,  2016, \mn@doi [\mnras] {10.1093/mnras/stw824}, \href
  {https://ui.adsabs.harvard.edu/abs/2016MNRAS.459.3460K} {459, 3460}

\bibitem[\protect\citeauthoryear{{Kruijssen} et~al.,}{{Kruijssen}
  et~al.}{2019}]{kruijssen2019}
{Kruijssen} J.~M.~D.,  et~al., 2019, \mn@doi [\nat]
  {10.1038/s41586-019-1194-3}, \href
  {https://ui.adsabs.harvard.edu/abs/2019Natur.569..519K} {569, 519}

\bibitem[\protect\citeauthoryear{{Krumholz}, {Dekel}  \& {McKee}}{{Krumholz}
  et~al.}{2012}]{krumholz2012}
{Krumholz} M.~R.,  {Dekel} A.,   {McKee} C.~F.,  2012, \mn@doi [\apj]
  {10.1088/0004-637X/745/1/69}, \href
  {https://ui.adsabs.harvard.edu/abs/2012ApJ...745...69K} {745, 69}

\bibitem[\protect\citeauthoryear{{Leitherer} et~al.,}{{Leitherer}
  et~al.}{1999}]{leitherer1999}
{Leitherer} C.,  et~al., 1999, \mn@doi [\apjs] {10.1086/313233}, \href
  {https://ui.adsabs.harvard.edu/abs/1999ApJS..123....3L} {123, 3}

\bibitem[\protect\citeauthoryear{{Leroy} et~al.,}{{Leroy}
  et~al.}{2013}]{leroy2013}
{Leroy} A.~K.,  et~al., 2013, \mn@doi [\aj] {10.1088/0004-6256/146/2/19}, \href
  {https://ui.adsabs.harvard.edu/abs/2013AJ....146...19L} {146, 19}

\bibitem[\protect\citeauthoryear{{Li}, {Ostriker}, {Cen}, {Bryan}  \&
  {Naab}}{{Li} et~al.}{2015}]{li2015}
{Li} M.,  {Ostriker} J.~P.,  {Cen} R.,  {Bryan} G.~L.,   {Naab} T.,  2015,
  \mn@doi [\apj] {10.1088/0004-637X/814/1/4}, \href
  {https://ui.adsabs.harvard.edu/abs/2015ApJ...814....4L} {814, 4}

\bibitem[\protect\citeauthoryear{{Liu}, {Tan}, {Marvil}, {Kong}, {Rosero},
  {Caselli}  \& {Cosentino}}{{Liu} et~al.}{2020}]{liu2020}
{Liu} M.,  {Tan} J.~C.,  {Marvil} J.,  {Kong} S.,  {Rosero} V.,  {Caselli} P.,
   {Cosentino} G.,  2020, arXiv e-prints, \href
  {https://ui.adsabs.harvard.edu/abs/2020arXiv201011294L} {p. arXiv:2010.11294}

\bibitem[\protect\citeauthoryear{{L{\'o}pez-Sepulcre}, {Codella}, {Cesaroni},
  {Marcelino}  \& {Walmsley}}{{L{\'o}pez-Sepulcre}
  et~al.}{2009}]{lopezsepulcre2009}
{L{\'o}pez-Sepulcre} A.,  {Codella} C.,  {Cesaroni} R.,  {Marcelino} N.,
  {Walmsley} C.~M.,  2009, \mn@doi [\aap] {10.1051/0004-6361/200912051}, \href
  {https://ui.adsabs.harvard.edu/abs/2009A&A...499..811L} {499, 811}

\bibitem[\protect\citeauthoryear{{L{\'o}pez-Sepulcre}
  et~al.,}{{L{\'o}pez-Sepulcre} et~al.}{2011}]{lopezsepulcre2011}
{L{\'o}pez-Sepulcre} A.,  et~al., 2011, \mn@doi [\aap]
  {10.1051/0004-6361/201015827}, \href
  {https://ui.adsabs.harvard.edu/abs/2011A&A...526L...2L} {526, L2}

\bibitem[\protect\citeauthoryear{{Marinacci}, {Sales}, {Vogelsberger}, {Torrey}
   \& {Springel}}{{Marinacci} et~al.}{2019}]{marinacci2019}
{Marinacci} F.,  {Sales} L.~V.,  {Vogelsberger} M.,  {Torrey} P.,   {Springel}
  V.,  2019, \mn@doi [\mnras] {10.1093/mnras/stz2391}, \href
  {https://ui.adsabs.harvard.edu/abs/2019MNRAS.489.4233M} {489, 4233}

\bibitem[\protect\citeauthoryear{{Martin-Pintado}, {Bachiller}  \&
  {Fuente}}{{Martin-Pintado} et~al.}{1992}]{martinpintado1992}
{Martin-Pintado} J.,  {Bachiller} R.,   {Fuente} A.,  1992, \aap, \href
  {https://ui.adsabs.harvard.edu/abs/1992A&A...254..315M} {254, 315}

\bibitem[\protect\citeauthoryear{{Martizzi}, {Faucher-Gigu{\`e}re}  \&
  {Quataert}}{{Martizzi} et~al.}{2015}]{martizzi2015}
{Martizzi} D.,  {Faucher-Gigu{\`e}re} C.-A.,   {Quataert} E.,  2015, \mn@doi
  [\mnras] {10.1093/mnras/stv562}, \href
  {https://ui.adsabs.harvard.edu/abs/2015MNRAS.450..504M} {450, 504}

\bibitem[\protect\citeauthoryear{{Martizzi}, {Fielding}, {Faucher-Gigu{\`e}re}
  \& {Quataert}}{{Martizzi} et~al.}{2016}]{martizzi2016}
{Martizzi} D.,  {Fielding} D.,  {Faucher-Gigu{\`e}re} C.-A.,   {Quataert} E.,
  2016, \mn@doi [\mnras] {10.1093/mnras/stw745}, \href
  {https://ui.adsabs.harvard.edu/abs/2016MNRAS.459.2311M} {459, 2311}

\bibitem[\protect\citeauthoryear{{Neufeld}, {Hollenbach}, {Kaufman}, {Snell},
  {Melnick}, {Bergin}  \& {Sonnentrucker}}{{Neufeld}
  et~al.}{2007}]{neufeld2007}
{Neufeld} D.~A.,  {Hollenbach} D.~J.,  {Kaufman} M.~J.,  {Snell} R.~L.,
  {Melnick} G.~J.,  {Bergin} E.~A.,   {Sonnentrucker} P.,  2007, \mn@doi [\apj]
  {10.1086/518857}, \href
  {https://ui.adsabs.harvard.edu/abs/2007ApJ...664..890N} {664, 890}

\bibitem[\protect\citeauthoryear{{Noriega-Crespo}, {Moro-Martin}, {Carey},
  {Morris}, {Padgett}, {Latter}  \& {Muzerolle}}{{Noriega-Crespo}
  et~al.}{2004}]{noriegacrespo2004}
{Noriega-Crespo} A.,  {Moro-Martin} A.,  {Carey} S.,  {Morris} P.~W.,
  {Padgett} D.~L.,  {Latter} W.~B.,   {Muzerolle} J.,  2004, \mn@doi [\apjs]
  {10.1086/423136}, \href
  {https://ui.adsabs.harvard.edu/abs/2004ApJS..154..402N} {154, 402}

\bibitem[\protect\citeauthoryear{{Padoan}, {Pan}, {Haugb{\o}lle}  \&
  {Nordlund}}{{Padoan} et~al.}{2016}]{padoan2016}
{Padoan} P.,  {Pan} L.,  {Haugb{\o}lle} T.,   {Nordlund} {\r{A}}.,  2016,
  \mn@doi [\apj] {10.3847/0004-637X/822/1/11}, \href
  {https://ui.adsabs.harvard.edu/abs/2016ApJ...822...11P} {822, 11}

\bibitem[\protect\citeauthoryear{{Padoan}, {Haugb{\o}lle}, {Nordlund}  \&
  {Frimann}}{{Padoan} et~al.}{2017}]{padoan2017}
{Padoan} P.,  {Haugb{\o}lle} T.,  {Nordlund} {\r{A}}.,   {Frimann} S.,  2017,
  \mn@doi [\apj] {10.3847/1538-4357/aa6afa}, \href
  {https://ui.adsabs.harvard.edu/abs/2017ApJ...840...48P} {840, 48}

\bibitem[\protect\citeauthoryear{{Parmentier}}{{Parmentier}}{2011}]{parmentier2011}
{Parmentier} G.,  2011, \mn@doi [\mnras] {10.1111/j.1365-2966.2011.18269.x},
  \href {https://ui.adsabs.harvard.edu/abs/2011MNRAS.413.1899P} {413, 1899}

\bibitem[\protect\citeauthoryear{{Petre}, {Szymkowiak}, {Seward}  \&
  {Willingale}}{{Petre} et~al.}{1988}]{petre1988}
{Petre} R.,  {Szymkowiak} A.~E.,  {Seward} F.~D.,   {Willingale} R.,  1988,
  \mn@doi [\apj] {10.1086/166922}, \href
  {https://ui.adsabs.harvard.edu/abs/1988ApJ...335..215P} {335, 215}

\bibitem[\protect\citeauthoryear{{Reach}, {Tram}, {Richter}, {Gusdorf}  \&
  {DeWitt}}{{Reach} et~al.}{2019}]{reach2019}
{Reach} W.~T.,  {Tram} L.~N.,  {Richter} M.,  {Gusdorf} A.,   {DeWitt} C.,
  2019, \mn@doi [\apj] {10.3847/1538-4357/ab41f7}, \href
  {https://ui.adsabs.harvard.edu/abs/2019ApJ...884...81R} {884, 81}

\bibitem[\protect\citeauthoryear{{Rho} \& {Petre}}{{Rho} \&
  {Petre}}{1998}]{rho1998}
{Rho} J.,  {Petre} R.,  1998, \mn@doi [\apjl] {10.1086/311538}, \href
  {https://ui.adsabs.harvard.edu/abs/1998ApJ...503L.167R} {503, L167}

\bibitem[\protect\citeauthoryear{{Sanderson} et~al.,}{{Sanderson}
  et~al.}{2018}]{sanderson2018}
{Sanderson} R.~E.,  et~al., 2018, \mn@doi [\apj] {10.3847/1538-4357/aaeb33},
  \href {https://ui.adsabs.harvard.edu/abs/2018ApJ...869...12S} {869, 12}

\bibitem[\protect\citeauthoryear{{Scannapieco}, {Tissera}, {White}  \&
  {Springel}}{{Scannapieco} et~al.}{2008}]{scannapieco2008}
{Scannapieco} C.,  {Tissera} P.~B.,  {White} S. D.~M.,   {Springel} V.,  2008,
  \mn@doi [\mnras] {10.1111/j.1365-2966.2008.13678.x}, \href
  {https://ui.adsabs.harvard.edu/abs/2008MNRAS.389.1137S} {389, 1137}

\bibitem[\protect\citeauthoryear{{Schilke}, {Walmsley}, {Pineau des Forets}  \&
  {Flower}}{{Schilke} et~al.}{1997}]{schilke1997}
{Schilke} P.,  {Walmsley} C.~M.,  {Pineau des Forets} G.,   {Flower} D.~R.,
  1997, \aap, \href {https://ui.adsabs.harvard.edu/abs/1997A&A...321..293S}
  {321, 293}

\bibitem[\protect\citeauthoryear{{Sedov}}{{Sedov}}{1959}]{sedov1959}
{Sedov} L.~I.,  1959, {Similarity and Dimensional Methods in Mechanics}

\bibitem[\protect\citeauthoryear{{Seifried}, {Haid}, {Walch}, {Borchert}  \&
  {Bisbas}}{{Seifried} et~al.}{2020}]{seifried2020}
{Seifried} D.,  {Haid} S.,  {Walch} S.,  {Borchert} E.~M.~A.,   {Bisbas} T.~G.,
   2020, \mn@doi [\mnras] {10.1093/mnras/stz3563}, \href
  {https://ui.adsabs.harvard.edu/abs/2020MNRAS.492.1465S} {492, 1465}

\bibitem[\protect\citeauthoryear{{Shima}, {Tasker}  \& {Habe}}{{Shima}
  et~al.}{2017}]{shima2017}
{Shima} K.,  {Tasker} E.~J.,   {Habe} A.,  2017, \mn@doi [\mnras]
  {10.1093/mnras/stw3279}, \href
  {https://ui.adsabs.harvard.edu/abs/2017MNRAS.467..512S} {467, 512}

\bibitem[\protect\citeauthoryear{{Slane}, {Bykov}, {Ellison}, {Dubner}  \&
  {Castro}}{{Slane} et~al.}{2016}]{slane2016}
{Slane} P.,  {Bykov} A.,  {Ellison} D.~C.,  {Dubner} G.,   {Castro} D.,  2016,
  {Supernova Remnants Interacting with Molecular Clouds: X-Ray and Gamma-Ray
  Signatures}.
p.~187, \mn@doi{10.1007/978-1-4939-3547-5\_6}

\bibitem[\protect\citeauthoryear{{Smith}, {Sijacki}  \& {Shen}}{{Smith}
  et~al.}{2018}]{smith2018}
{Smith} M.~C.,  {Sijacki} D.,   {Shen} S.,  2018, \mn@doi [\mnras]
  {10.1093/mnras/sty994}, \href
  {https://ui.adsabs.harvard.edu/abs/2018MNRAS.478..302S} {478, 302}

\bibitem[\protect\citeauthoryear{{Sobolev}}{{Sobolev}}{1957}]{sobolev1957}
{Sobolev} V.~V.,  1957, \sovast, \href
  {https://ui.adsabs.harvard.edu/abs/1957SvA.....1..678S} {1, 678}

\bibitem[\protect\citeauthoryear{{Stephens} et~al.,}{{Stephens}
  et~al.}{2017}]{stephens2017}
{Stephens} I.~W.,  et~al., 2017, \mn@doi [\apj] {10.3847/1538-4357/aa8262},
  \href {https://ui.adsabs.harvard.edu/abs/2017ApJ...846...16S} {846, 16}

\bibitem[\protect\citeauthoryear{{Stinson}, {Seth}, {Katz}, {Wadsley},
  {Governato}  \& {Quinn}}{{Stinson} et~al.}{2006}]{stinson2006}
{Stinson} G.,  {Seth} A.,  {Katz} N.,  {Wadsley} J.,  {Governato} F.,   {Quinn}
  T.,  2006, \mn@doi [\mnras] {10.1111/j.1365-2966.2006.11097.x}, \href
  {https://ui.adsabs.harvard.edu/abs/2006MNRAS.373.1074S} {373, 1074}

\bibitem[\protect\citeauthoryear{{Taylor}}{{Taylor}}{1950}]{taylor1950}
{Taylor} G.,  1950, \mn@doi [Proceedings of the Royal Society of London Series
  A] {10.1098/rspa.1950.0049}, \href
  {https://ui.adsabs.harvard.edu/abs/1950RSPSA.201..159T} {201, 159}

\bibitem[\protect\citeauthoryear{{Teyssier}, {Pontzen}, {Dubois}  \&
  {Read}}{{Teyssier} et~al.}{2013}]{teyssier2013}
{Teyssier} R.,  {Pontzen} A.,  {Dubois} Y.,   {Read} J.~I.,  2013, \mn@doi
  [\mnras] {10.1093/mnras/sts563}, \href
  {https://ui.adsabs.harvard.edu/abs/2013MNRAS.429.3068T} {429, 3068}

\bibitem[\protect\citeauthoryear{{Troja}, {Bocchino}  \& {Reale}}{{Troja}
  et~al.}{2006}]{troja2006}
{Troja} E.,  {Bocchino} F.,   {Reale} F.,  2006, \mn@doi [\apj]
  {10.1086/506378}, \href
  {https://ui.adsabs.harvard.edu/abs/2006ApJ...649..258T} {649, 258}

\bibitem[\protect\citeauthoryear{{Troja}, {Bocchino}, {Miceli}  \&
  {Reale}}{{Troja} et~al.}{2008}]{troja2008}
{Troja} E.,  {Bocchino} F.,  {Miceli} M.,   {Reale} F.,  2008, \mn@doi [\aap]
  {10.1051/0004-6361:20079123}, \href
  {https://ui.adsabs.harvard.edu/abs/2008A&A...485..777T} {485, 777}

\bibitem[\protect\citeauthoryear{{Ustamujic}, {Orlando}, {Greco}, {Miceli},
  {Bocchino}, {Tutone}  \& {Peres}}{{Ustamujic} et~al.}{2021}]{ustamujic2021}
{Ustamujic} S.,  {Orlando} S.,  {Greco} E.,  {Miceli} M.,  {Bocchino} F.,
  {Tutone} A.,   {Peres} G.,  2021, \mn@doi [\aap]
  {10.1051/0004-6361/202039940}, \href
  {https://ui.adsabs.harvard.edu/abs/2021A&A...649A..14U} {649, A14}

\bibitem[\protect\citeauthoryear{{Vasyunina}, {Linz}, {Henning}, {Zinchenko},
  {Beuther}  \& {Voronkov}}{{Vasyunina} et~al.}{2011}]{vasyunina2011}
{Vasyunina} T.,  {Linz} H.,  {Henning} T.,  {Zinchenko} I.,  {Beuther} H.,
  {Voronkov} M.,  2011, \mn@doi [\aap] {10.1051/0004-6361/201014974}, \href
  {https://ui.adsabs.harvard.edu/abs/2011A&A...527A..88V} {527, A88}

\bibitem[\protect\citeauthoryear{{Vaupre}}{{Vaupre}}{2015}]{vaupre2015}
{Vaupre} S.,  2015, PhD thesis, University of Grenoble

\bibitem[\protect\citeauthoryear{{Verheyen}, {Messineo}  \&
  {Menten}}{{Verheyen} et~al.}{2012}]{verheyen2012}
{Verheyen} L.,  {Messineo} M.,   {Menten} K.~M.,  2012, \mn@doi [\aap]
  {10.1051/0004-6361/201118265}, \href
  {https://ui.adsabs.harvard.edu/abs/2012A&A...541A..36V} {541, A36}

\bibitem[\protect\citeauthoryear{{Wetzel}, {Hopkins}, {Kim},
  {Faucher-Gigu{\`e}re}, {Kere{\v{s}}}  \& {Quataert}}{{Wetzel}
  et~al.}{2016}]{wetzel2016}
{Wetzel} A.~R.,  {Hopkins} P.~F.,  {Kim} J.-h.,  {Faucher-Gigu{\`e}re} C.-A.,
  {Kere{\v{s}}} D.,   {Quataert} E.,  2016, \mn@doi [\apjl]
  {10.3847/2041-8205/827/2/L23}, \href
  {https://ui.adsabs.harvard.edu/abs/2016ApJ...827L..23W} {827, L23}

\bibitem[\protect\citeauthoryear{{White}, {Rainey}, {Hayashi}  \&
  {Kaifu}}{{White} et~al.}{1987}]{white1987}
{White} G.~J.,  {Rainey} R.,  {Hayashi} S.~S.,   {Kaifu} N.,  1987, \aap, \href
  {https://ui.adsabs.harvard.edu/abs/1987A&A...173..337W} {173, 337}

\bibitem[\protect\citeauthoryear{{Wootten}}{{Wootten}}{1977}]{wootten1977}
{Wootten} H.~A.,  1977, \mn@doi [\apj] {10.1086/155485}, \href
  {https://ui.adsabs.harvard.edu/abs/1977ApJ...216..440W} {216, 440}

\bibitem[\protect\citeauthoryear{{Wright} et~al.,}{{Wright}
  et~al.}{2010}]{wright2010}
{Wright} E.~L.,  et~al., 2010, \mn@doi [\aj] {10.1088/0004-6256/140/6/1868},
  \href {https://ui.adsabs.harvard.edu/abs/2010AJ....140.1868W} {140, 1868}

\bibitem[\protect\citeauthoryear{{Zeng} et~al.,}{{Zeng}
  et~al.}{2017}]{zeng2017}
{Zeng} S.,  et~al., 2017, \mn@doi [\aap] {10.1051/0004-6361/201630210}, \href
  {https://ui.adsabs.harvard.edu/abs/2017A&A...603A..22Z} {603, A22}

\bibitem[\protect\citeauthoryear{{Zhang} \& {Chevalier}}{{Zhang} \&
  {Chevalier}}{2019}]{zhangchevalier2019}
{Zhang} D.,  {Chevalier} R.~A.,  2019, \mn@doi [\mnras]
  {10.1093/mnras/sty2769}, \href
  {https://ui.adsabs.harvard.edu/abs/2019MNRAS.482.1602Z} {482, 1602}

\bibitem[\protect\citeauthoryear{{Zhang}, {Hunter}, {Brand}, {Sridharan},
  {Cesaroni}, {Molinari}, {Wang}  \& {Kramer}}{{Zhang}
  et~al.}{2005}]{zhang2005}
{Zhang} Q.,  {Hunter} T.~R.,  {Brand} J.,  {Sridharan} T.~K.,  {Cesaroni} R.,
  {Molinari} S.,  {Wang} J.,   {Kramer} M.,  2005, \mn@doi [\apj]
  {10.1086/429660}, \href
  {https://ui.adsabs.harvard.edu/abs/2005ApJ...625..864Z} {625, 864}

\bibitem[\protect\citeauthoryear{{Zhang}, {Gao}  \& {Wang}}{{Zhang}
  et~al.}{2010}]{zhang2010}
{Zhang} Z.,  {Gao} Y.,   {Wang} J.,  2010, \mn@doi [Science China Physics,
  Mechanics, and Astronomy] {10.1007/s11433-010-4010-5}, \href
  {https://ui.adsabs.harvard.edu/abs/2010SCPMA..53.1357Z} {53, 1357}

\bibitem[\protect\citeauthoryear{{Ziurys}, {Snell}  \& {Dickman}}{{Ziurys}
  et~al.}{1989}]{ziurys1989}
{Ziurys} L.~M.,  {Snell} R.~L.,   {Dickman} R.~L.,  1989, \mn@doi [\apj]
  {10.1086/167544}, \href
  {https://ui.adsabs.harvard.edu/abs/1989ApJ...341..857Z} {341, 857}

\bibitem[\protect\citeauthoryear{{van Dishoeck}, {Jansen}  \& {Phillips}}{{van
  Dishoeck} et~al.}{1993}]{vanDishoeck1993}
{van Dishoeck} E.~F.,  {Jansen} D.~J.,   {Phillips} T.~G.,  1993, \aap, \href
  {https://ui.adsabs.harvard.edu/abs/1993A&A...279..541V} {279, 541}

\bibitem[\protect\citeauthoryear{{van der Tak}, {Black}, {Sch{\"o}ier},
  {Jansen}  \& {van Dishoeck}}{{van der Tak} et~al.}{2007}]{tak2007}
{van der Tak} F.~F.~S.,  {Black} J.~H.,  {Sch{\"o}ier} F.~L.,  {Jansen} D.~J.,
   {van Dishoeck} E.~F.,  2007, \mn@doi [\aap] {10.1051/0004-6361:20066820},
  \href {https://ui.adsabs.harvard.edu/abs/2007A&A...468..627V} {468, 627}

\makeatother
\end{thebibliography}


% Don't change these lines
\bsp	% typesetting comment
\label{lastpage}
\end{document}